\documentclass[
journal=jpcbfk, 
manuscript=article,
layout=twocolumn   
]{achemso}

\usepackage[T1]{fontenc} 

\usepackage{graphicx}
\graphicspath{ {./Figures4Submission/} }

\usepackage{pbox}
\usepackage{tabularx}

\usepackage{mathtools} 
\usepackage{amssymb} 
\usepackage[flushleft]{threeparttable}

\usepackage{dcolumn}
\usepackage{bm}
\usepackage[normalem]{ulem}
\usepackage{xcolor}
\usepackage{hyperref}

\definecolor{OliveGreen}{rgb}{0,0.6,0}
\definecolor{auburn}{rgb}{0.43, 0.21, 0.1}
\definecolor{BlueViolet}{rgb}{0.54, 0.17, 0.89}
\definecolor{HokieOrange}{RGB}{232, 119, 34}
\definecolor{HokieMaroon}{RGB}{134, 31, 65}
\definecolor{Black}{RGB}{0, 0, 0}

\usepackage{appendix}
\usepackage{chngcntr}
\usepackage{etoolbox}
\usepackage{lipsum}
\usepackage{makecell}
\usepackage{natbib}
\usepackage{subcaption}

\author{Jared McDonald}
\affiliation{Materials Science \& Engineering Department, Virginia Tech, Blacksburg, VA 24061, USA}
\email{jmcdonald@vt.edu}

\author{Michael R. von Spakovsky}
\affiliation{Mechanical Engineering Department, Virginia Tech, Blacksburg, VA 24061, USA}
\email{vonspako@vt.edu}

\author{William T. Reynolds Jr.}
\affiliation{Materials Science \& Engineering Department, Virginia Tech, Blacksburg, VA 24061, USA}
\email{reynolds@vt.edu}

\title{Predicting Polymer Brush Behavior in Solvents using the Steepest-Entropy-Ascent Quantum Thermodynamic Framework}

\begin{document}



\begin{tocentry}
\includegraphics[height=0.18\textheight]{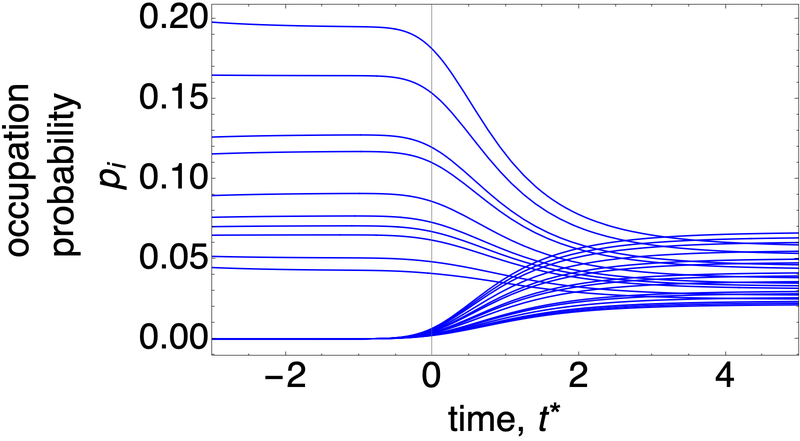}
\end{tocentry}

\begin{abstract}
The steepest-entropy-ascent quantum thermodynamic (SEAQT) framework is utilized to study the effects of temperature on polymer brushes. The brushes are represented by a discrete energy spectrum and energy degeneracies obtained through the Replica-Exchange Wang-Landau algorithm. The SEAQT equation of motion is applied to the density of states to establish a unique kinetic path from an initial thermodynamic state to a stable equilibrium state. The kinetic path describes the brush's evolution in state space as it interacts with a thermal reservoir. The predicted occupation probabilities along the kinetic path are used to determine expected thermodynamic and structural properties. The polymer density profile of a polystyrene brush in cyclohexane solvent is predicted using the equation of motion, and it agrees qualitatively with experimental density profiles. The Flory-Huggins parameter chosen to describe brush-solvent interactions affects the solvent distribution in the brush but has minimal impact on the polymer density profile. Three types of non-equilibrium kinetic paths with differing amounts of entropy production are considered: a heating path, a cooling path, and a heating-cooling path. Properties such as tortuosity, radius of gyration, brush density, solvent density, and brush chain conformations are calculated for each path.
\end{abstract}

\maketitle

\section{Introduction}

Polymer brushes are a versatile class of coatings where one end of the polymer chains is grafted to a surface, with end-grafted polymer chains having numerous applications in colloid dispersion, anti-fouling surfaces, lubrication, adhesion, and sensing~\cite{Ayres2010, Buhl2020, LiDanyang2021,Chen2017PolymerBrushesPerspective}. Experimental studies have investigated the behavior of neutral and electrostatically charged chain groups interacting with polar and nonpolar solvents and gases~\cite{Karim1994, Field1992, Borisov1997, Biesalski1999, Biesalski2000, Galvin2014,vanEck2020, Orski2015, Galvin2016, Sun2016, Karim1994}. Computational simulations using molecular dynamics and Monte Carlo methods have explored the near-equilibrium molecular conformations of brushes in various chemical environments~\cite{vanEck2020, Jentzsch2014, Dimitrov2007, Galuschko2019, Lai1991, Auroy1991, Lai1992, Auroy1991, Laradji1994}.

Acquiring information on the dynamic evolution of brushes in response to temperature changes has proven challenging. Neutron reflectometry has been utilized to measure polymer density as a function of distance from the grafting wall~\cite{Karim1994, Field1992}; however, this technique requires temperature equilibration of samples for extended periods ranging from minutes to hours~\cite{Russell1990}. Therefore, currently available experimental data is insensitive to the polymer chain dynamics on shorter time scales.

Extracting their dynamic evolution from computational simulations is also challenging. In molecular dynamics and Monte Carlo methods, a stable brush configuration at a chosen temperature is obtained from an arbitrary initial arrangement~\cite{Lai1991, Dimitrov2007, He2007} by minimizing free energy. Repeating this procedure at different temperatures produces an idealized, quasi-equilibrium path suitable for systems at stable or perhaps even metastable equilibrium. The latter can occur since minimizing the energy through chain relaxations can lead to local energy minima (as opposed to the global minimum) as a result of becoming trapped in such minima by energy barriers between chain conformations~\cite{Lai1991}, leading to a cessation of chain evolution. To overcome this latter issue, one must resort to ``averaging'' structural parameters over many simulations with different starting points, which introduces stochastic uncertainty. Moreover, molecular dynamics may require unrealistically high simulation temperatures to generate the conformational changes necessary to reach low-energy configurations in reasonable times. Additionally, some Monte Carlo methods may not be able to determine relaxation times when the minimum energy conformation cannot be reached along a sequence of successive chain movements~\cite{Mavrantzas2021}. These movement paths cannot be produced by diffusive motion, so relaxation times cannot be estimated for them. Although there are computational models of non-equilibrium brush dynamics available for the shearing conditions relevant to lubrication and adhesion applications~\cite{Binder2011, Milchev2010, Baker2000}, these approaches cannot be generalized to simple thermal systems.

In this contribution, we present a modeling scheme to investigate {\em non-equilibrium} properties of polymer brushes during heating or cooling. To achieve this, we use a Monte Carlo method to determine an energy eigenstructure that serves as input to a kinetic equation of motion derived from a quantum-inspired, non-conservative (dissipative) framework. The energy eigenstructure lists all the possible internal energy levels that a discretized thermodynamic system can access together with the degeneracies of these energy levels. The equation of motion, on the other hand, employs a variational principle to determine how the occupancies of the individual energy levels evolve over time as the system moves from an initial non-equilibrium state to a stable equilibrium state.

The energy eigenstructure of the polymer brush -- also called the ``density of states'' in some contexts -- characterizes how the discretized energy levels and their corresponding degeneracies vary with system parameters such as chain length, solvent type, and concentration. To obtain the degeneracy of the energy levels, we employ the Replica-Exchange Wang-Landau algorithm~\cite{Li2014, Vogel2013, WangLandau2001a, WangLandau2001b}, a parallelized, entropic, and non-Markovian Monte-Carlo sampling technique. This algorithm has been extensively utilized for polymer folding systems, providing estimates of equilibrium properties such as heat capacity, radius of gyration, and polymer chain tortuosity~\cite{Wust2012, Farris2019, Taylor2020Confine, Taylor2020Crowd}. However, since an equilibrium approach cannot account for non-equilibrium effects during dynamic changes like heating or cooling, canonical distributions applied to an energy eigenstructure generated with the Replica-Exchange Wang-Landau algorithm alone cannot accurately predict dynamic evolution since this always involves non-equilibrium transition states~\cite{Wang2011, Wust2012, Farris2019}.

To treat non-equilibrium processes explicitly, this work applies the steepest-entropy-ascent quantum thermodynamic (SEAQT) framework~\cite{Li2016a,Li2016b,Li2016c,Li2018,Li2018steepest,Li2017study, Li2018multiscale,McDonald2022capillary,Yamada2018method,Yamada2019,Yamada2019kineticpartI,Yamada2019spin,Yamada2020kineticpartII,jhon2020,Cano2015steepest,kusaba2019,kusaba2017,Goswami2021,vonSpakovsky2020,Beretta2014steepest,Beretta2006,Beretta1984,Beretta1985,Hatsopoulos1976-I,Hatsopoulos1976-IIa,Hatsopoulos1976-IIb,Hatsopoulos1976-III,Younis2022,McDonald2023polymer} to determine a brush system's kinetic path from an arbitrary initial state to stable equilibrium. The framework uses a postulated equation of motion to calculate how energy redistributes among a brush's discrete energy levels as it evolves through transition states to equilibrium.  The equation of motion relies only upon the system's energy levels and degeneracies and the principle of steepest entropy ascent. Solving the equation of motion provides the time-dependent occupation probabilities for the energy levels; this is a full description of the kinetic path. The SEAQT framework is distinct from conventional kinetic descriptions in that the SEAQT equation of motion predicts the state evolution without resorting to any {\em a priori} assumptions about the physical or kinetic mechanisms underlying the physical processes. 

It is important to note here is that for systems not at equilibrium, energy minimization alone is insufficient to predict a realistic (i.e., non-equilibrium) path that is consistent with the system’s underlying equation of state; an additional postulate is needed. The one used here is the steepest-entropy-ascent principle on which our equation of motion is based. Given an arbitrary initial state, this equation predicts a unique irreversible path that ends at stable equilibrium. It cannot get trapped at local energy minima. Our procedure is more involved than existing approaches for two reasons. The first is that the entropy must be determined and this requires two steps, namely, i) knowing the system energy levels and degeneracies, which can be determined using Wang-Landau as is done here (or some other approach, e.g., a multicanonical approach) and ii) knowing how the energy is continually redistributed among the energy levels along the non-equilibrium path. This second step requires a general equation of motion such as the one based on steepest-entropy-ascent quantum thermodynamics and such an equation is not present in current approaches. The second reason that our approach is more involved is that  connecting the series of non-equilibrium states predicted by the SEAQT equation of motion to physical properties requires an additional procedure or step, which is described below. Clearly, these additional steps are not necessary if one only cares about equilibrium properties, but they are unavoidable when predicting brush properties along a non-equilibrium (i.e., irreversible) path.

Now, an essential question to address at the outset is the following: how fundamental is the steepest-entropy-ascent principle?  Beretta~\cite{Beretta2020} has argued that a large body of research over the past forty years on the essential features of non-equilibrium natural phenomena qualify the principle as a `great law of nature,' worthy of designation as a fourth law of thermodynamics.  It is a variational principle~\cite{Beretta2014steepest} consistent with the other laws of thermodynamics and the laws of mechanics. Although it has been called by other names and expressed in different equivalent forms, steepest entropy ascent has its roots in Ziegler's maximum-entropy-production principle from the early 1960's~\cite{Ziegler1963some}. Experimental evidence suggests natural processes maximize entropy production at each instant of time~\cite{Yamada2020kineticpartII, Martyushev2013,Martyushev2021}, and as shown, for example, by Li and von Spakovsky~\cite{Li2018steepest, Li2016b},  the steepest-entropy-ascent principle correctly predicts expected temperature and concentration profiles under steady state conditions without having to assume an underlying transport law. In terms of the range of its applicability, both linear and nonlinear non-equilibrium thermodynamics can be deduced from the maximum-entropy-production principle~\cite{Ziegler1957thermodynamik, Ziegler1963some, Ziegler1983chemical, Ziegler1983introduction, Ziegler1987principle, Beretta1984, Li2016b, Li2018}. In fact, as Beretta \cite{Beretta2020} has shown, the nonlinear extensions of the Onsager formalism to the far-from-equilibrium region can be derived directly from the steepest-entropy-ascent rate-controlled constrained-equilibrium formalism as can the dispersion-dissipation relations. A similar extension of the Onsager formalism to the far-from-equilibrium region using the steepest-entropy-ascent hypo-equilibrium formalism has been derived by Li and von Spakovsky \cite{Li2018}. While the steepest-entropy-ascent principle is treated here as a postulate, it has a long history, has a firm mathematical basis~\cite{BerettaPhD1981, Beretta1984, Beretta1985, Hatsopoulos1976-I, Hatsopoulos1976-IIa, Hatsopoulos1976-IIb, Hatsopoulos1976-III, Gyftopoulos1997}, and has been tested in a wide range of quantum and classical applications~\cite{Li2016a,Li2016b,Li2016c,Li2018,Li2018steepest,Li2018steepest,Li2017study, Li2018multiscale,Yamada2018method,Yamada2019,Yamada2019kineticpartI,Yamada2019spin,Yamada2020kineticpartII,jhon2020,Cano2015steepest,kusaba2019,kusaba2017,vonSpakovsky2020,McDonald2022capillary,McDonald2023polymer,Younis2022}. Additional discussion of the generality of the steepest-entropy-ascent principle, can be found in references:~\cite{McDonald2022capillary, McDonald2023polymer, Beretta2020,Beretta2014steepest}.

The present paper is structured as follows: Section~\nameref{subsec:EnergeticModel} outlines the energetic model of a polymer brush in a solvent, Section~\nameref{subsec:WangLandau} details the construction of the energy eigenstructure for the brush system, Section~\nameref{subsec:SEAQT-EOM} provides a summary of the basis for the SEAQT equation of motion, and Section~\nameref{sec:MicroLink} links the kinetic path in state space to physical microstructures. The results obtained from applying the SEAQT framework to a styrene brush in both cyclohexane and toluene solvents are presented in Section~\nameref{sec:results}.

\section{Method}

\subsection{Energetic Model \label{subsec:EnergeticModel}} 

The polymer brush can be described in terms of the extensive energy and entropy of a non-isolated thermodynamic system. The energy of the system is calculated as the sum of interactions between pairs of molecules. Each interaction is represented by a pair-potential function that depends upon the distance between the interacting molecules. The distance between the pairs is generated using the bond fluctuation model on an underlying lattice; this approach has been described earlier by others~\cite{Carmesin1988, Deutsch1991} and employed with Monte Carlo models applied to polymer brushes~\cite{Chen2002,Jentzsch2013,Jentzsch2014,Binder2011,Kopf1996,Lai1992Bidip,Lai1991}.

The three-dimensional lattice is constructed from two interpenetrating cubic sub-lattices that are shifted from each other by half a lattice parameter in each of the three Cartesian directions. The combined sub-lattices are equivalent to a body-centered cubic Bravais lattice. The bond fluctuation approach combines the computational efficiency of a lattice model with the ability to describe branched polymer chains with variable monomer bond lengths, while also maintaining an excluded volume around each occupied monomer site~\cite{Carmesin1988, Deutsch1991}.

\begin{figure}
	\begin{center}
		\includegraphics[width=0.45\textwidth]{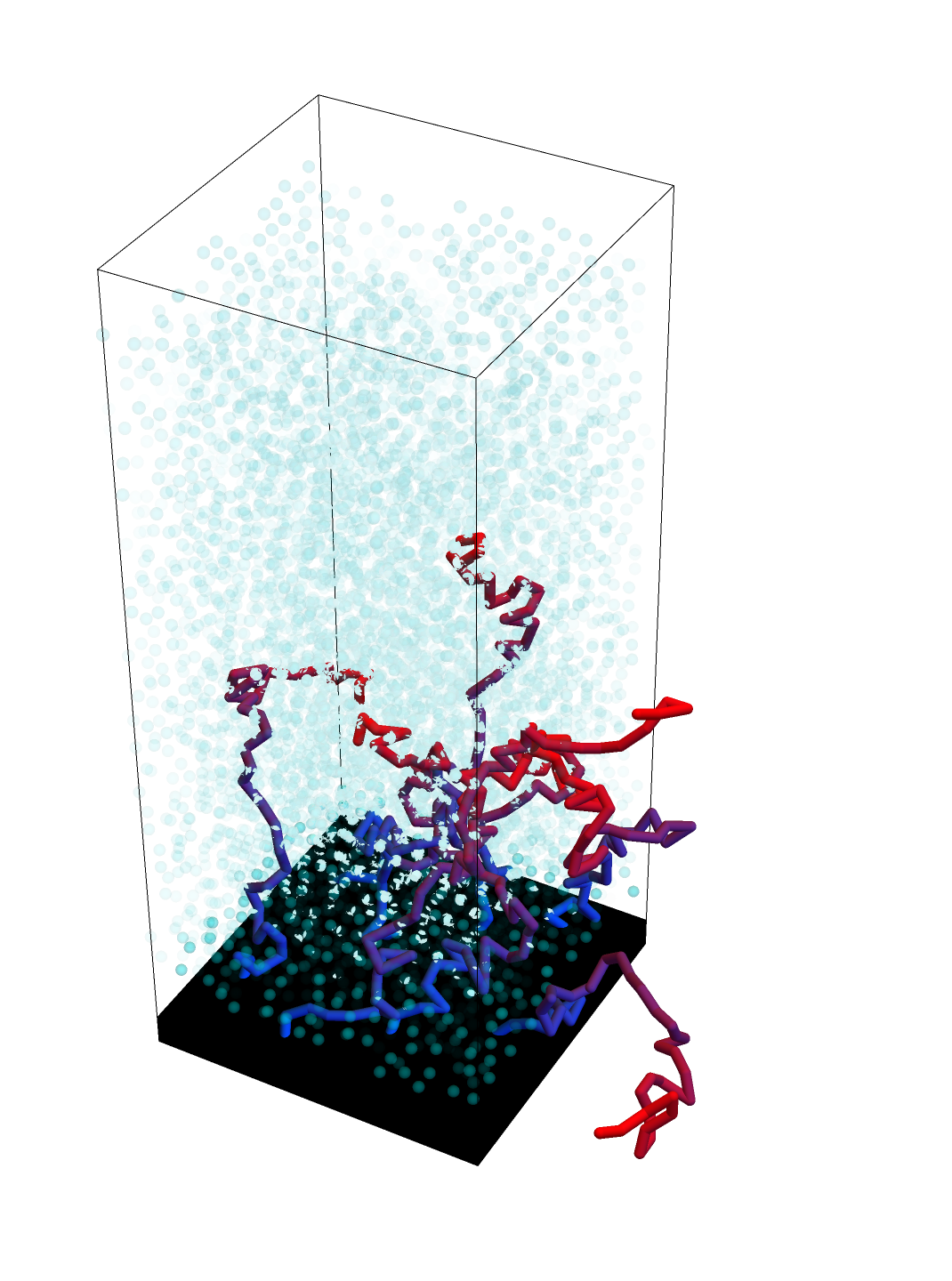}
		\caption{ Example conformation of a simulated styrene brush in a solvent. The simulation volume is $34\times34\times80$ sublattice spacings in size with 7500 solvent molecules (cyan). The grafting plane is the solid polyhedron below the simulation volume, and there are 8 styrene chains each of which is 40 monomers long (blue at their grafted ends and red at their free ends). The boundaries of the simulation volume are delineated by the solvent molecules. Because periodic boundary conditions are employed, the styrene chains are contained entirely within the simulation volume, but they are plotted in the figure extending beyond the boundaries of the simulation volume to avoid clutter and make the brush configuration more evident.}
		\label{ExampleBrush}
	\end{center}
\end{figure}

Figure~\ref{ExampleBrush} shows an example brush configuration within the simulation domain. The domain consists of a $34\times34\times80$ lattice with $17\times17\times40$ lattice parameters on the body-centered cubic lattice and with the grafting surface located at the bottom of the figure at $z=0$. Periodic boundary conditions are applied in the vertical planes of the four lateral sides of the simulation block.

The brush itself is composed of eight end-grafted chains, each 40 monomers in length, and is surrounded by 7500 nonpolar solvent molecules (either cyclohexane or toluene) represented by the cyan balls in Figure~\ref{ExampleBrush}. Each polymer chain is colored with a blue-to-red gradient to aid in distinguishing chain conformations (blue at the grafted ends, red at the free ends). With this geometry, the bond fluctuation model allows for pair-potential interaction distances that are the following six multiples of the lattice parameter: $a\sqrt{4}$, $a\sqrt{5}$, $a\sqrt{6}$, $a\sqrt{8}$, $a\sqrt{9}$, and $a\sqrt{10}$. The final value represents the effective pair-potential cutoff distance. For the sake of expressing the properties of subsequent sections in familiar units, a ``mole'' of the simulated brush is taken to mean Avogadro's number of the simulation volume's formula unit: 320 styrene monomers and the 7500 solvent molecules.

In many lattice-based Monte Carlo simulations of end-grafted polymers, a simple energetic model is often used, whereby all interactions within the maximum interaction distance increment the energy by a constant~\cite{Lai1991, Jentzsch2013, Jentzsch2014}. However, in this work, a more elaborate version of the Kermer Grest molecular energy model~\cite{Grest1986, Dimitrov2007, He2007} is employed with a different potential function for each interaction. The potential functions depend upon the type of molecules in the interacting pairs. This approach allows for a more accurate representation of the interaction energies between different types of molecules and can lead to more realistic simulation results.

The discrete energy of a given configuration can be described by its eigenenergy or energy eigenlevel, denoted as $E_j$. This eigenenergy is computed as a summation over all pairwise energies of the molecules present in the configuration, which can be expressed as:

\begin{equation}
E_j = \frac{1}{2}\sum_{n=1}^{N_{\mathrm{tot}}} \left( \mathop{\sum_{m=1}}_{m\neq n}^{N_{\mathrm{tot}}} V_{n,m}^{\phi} \right) \;\;\;\; .
\label{eqn:Ej}
\end{equation}

Here, $m$ and $n$ represent two molecules that are interacting with each other, and the summations are over $N_{\text{tot}}$, the total number of molecules in the system. The 1/2 in this equation accounts for the pairing. The $N_{\text{tot}}$ parameter includes the molecules in the polymer chains, the molecules in the grafting plane, and the solvent molecules, i.e., $N_{\text{tot}} = N_{\text{monomers}} + N_{\text{wall}}  + N_{\text{solvent}}$.  The $V^\phi$ terms represent pair-potential functions that  depend upon the two molecules in the $m$ and $n$ pair. These functions are listed in Table~\ref{PairPotentials}. For example, when $m$ and $n$ are both monomers along a single polymer chain, $V^{\phi}$ becomes $V_{\text{FENE}}$; when $m$ and $n$ are both solvent molecules, $V^{\phi}$ becomes $V_{\text{ss}}$; when one of $m$ and $n$ is a monomer in a chain and the other is a solvent molecule, $V^{\phi}$ becomes $V_{\text{ps}}$; and so on. A practical benefit of using different potentials for different types of interactions is that different polymer-solvent systems can be treated by simply changing the parameters for the relevant interactions without having to fit all intermolecular interactions to a single average potential~\cite{Jentzsch2014}.

\begin{figure}
	\begin{center}
  \begin{subfigure}{0.38\textwidth}
        \includegraphics[width=\linewidth]{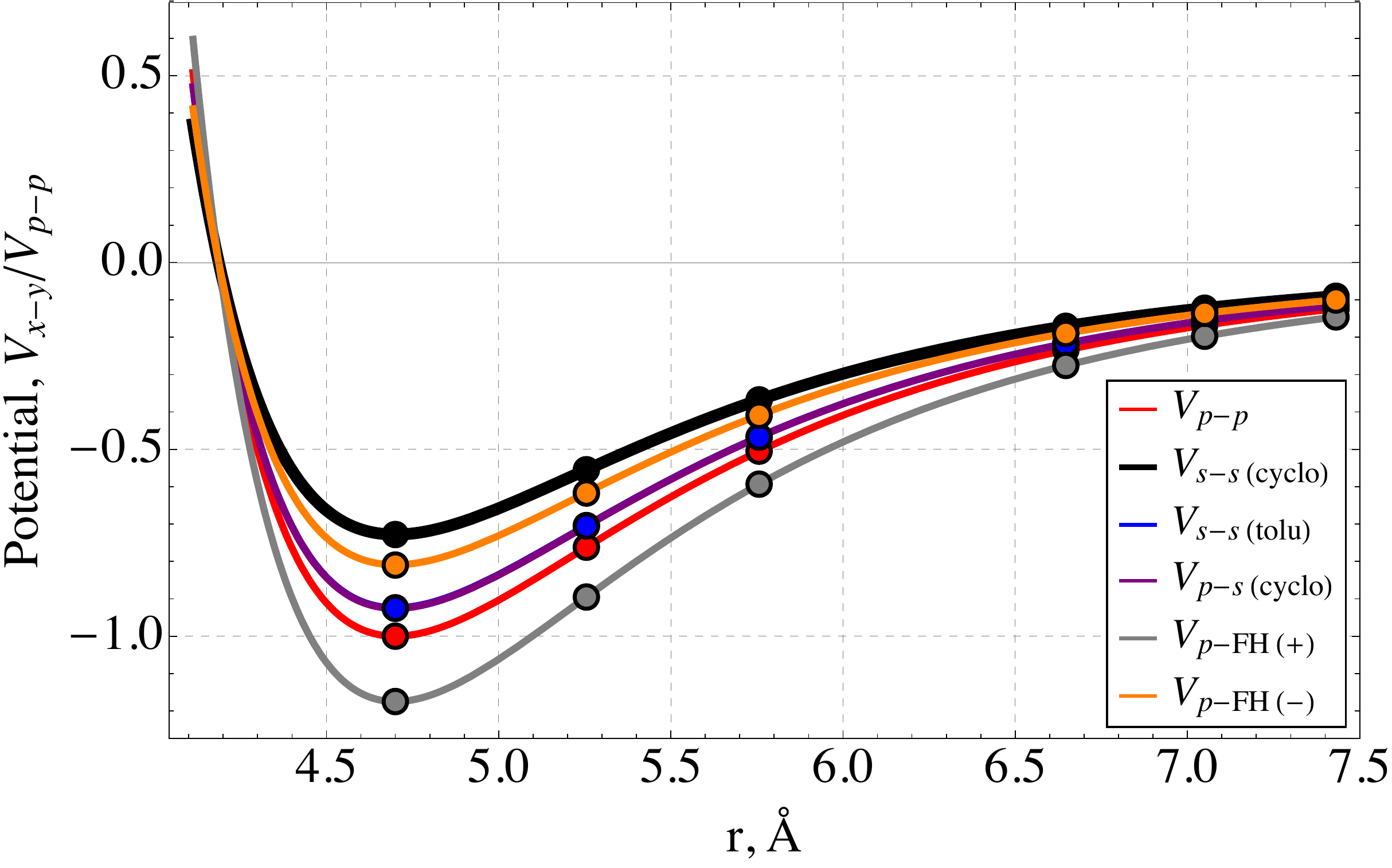}
        \caption{\vspace*{0mm}} 
        \label{PotentialsA}
%
        \includegraphics[width=\linewidth]{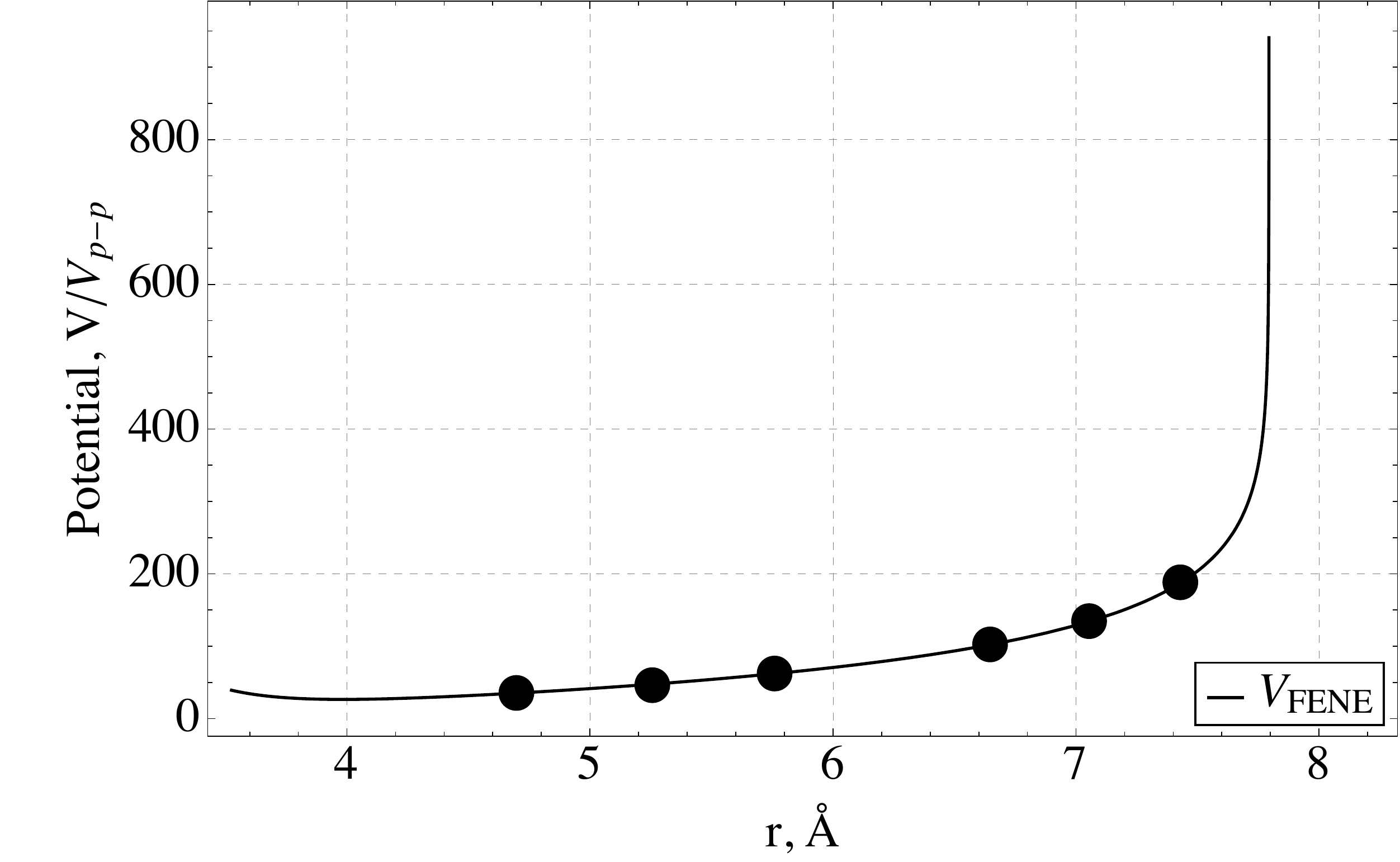}
        \caption{\vspace*{0mm}} 
        \label{PotentialsB}
%
        \includegraphics[width=\linewidth]{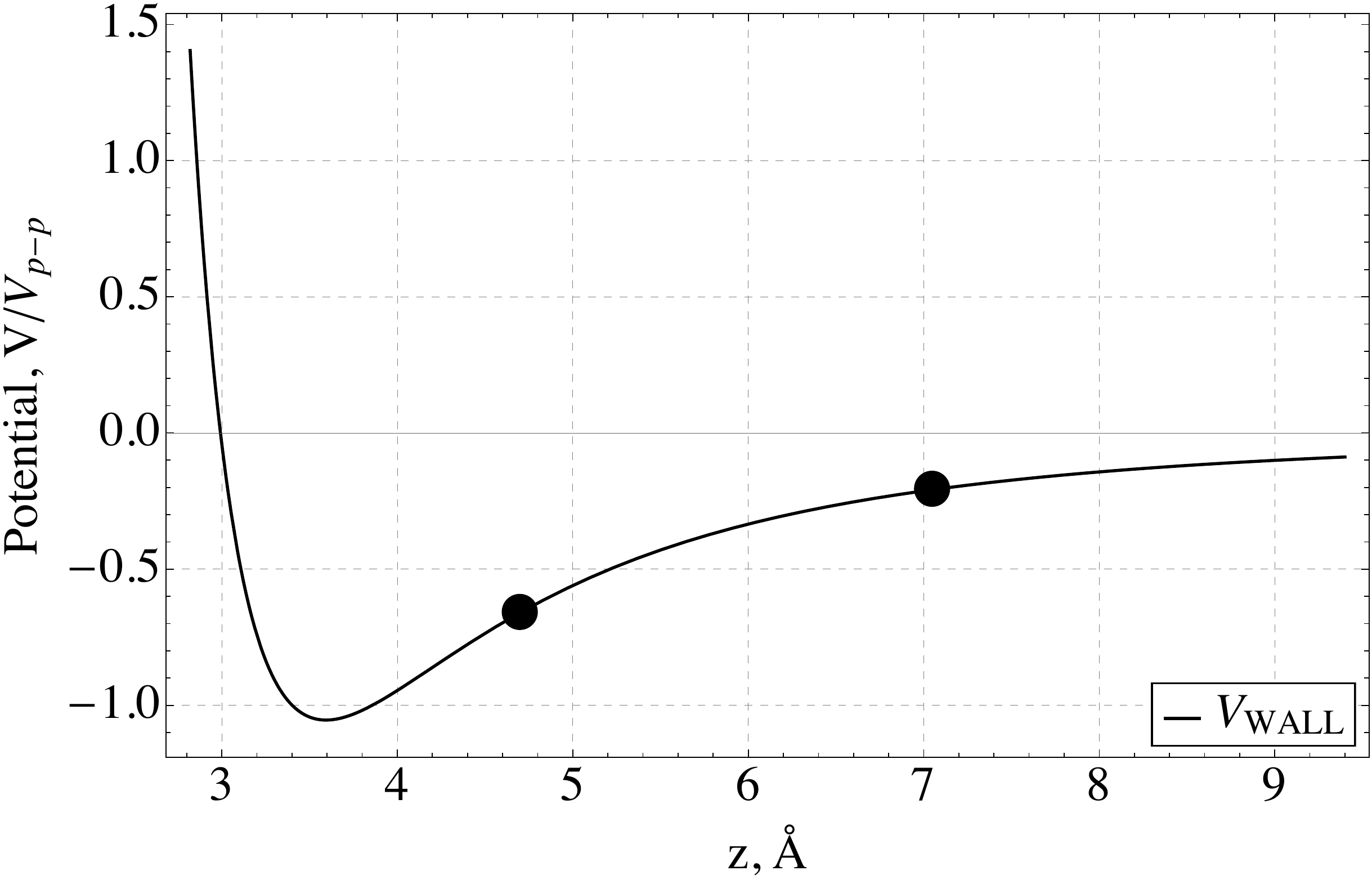}
        \caption{} \label{PotentialsC}
  \end{subfigure}
  \hspace*{\fill}
		\caption{\footnotesize The set of pair-potentials used in the simulations: (a) the 6-12 Lennard-Jones potentials representing non-bonding interactions (styrene--styrene molecules from different chains, solvent--solvent molecules, and styrene--solvent molecules; two cases for the styrene-toluene potential were estimated from positive and negative values of the Flory-Huggins parameter), (b) the FENE potential for interactions between monomers along a single styrene chain, and (c) the 9-3 Lennard-Jones potential between the grafting wall and all of the constituents. The points in the figure represent the interaction energies corresponding to the six allowed interaction distances of the bond fluctuation model with lattice parameter, $a=2.35$ \AA. The furthest distance, $7.4$ \AA, is the effective pair-potential cutoff distance. Because the grafting wall is a plane below the brush chains, there are only two possible interaction distances between the monomers in the chain and the wall molecules. }
		\label{Potentials}
	\end{center}
\end{figure}

The numerical value for the lattice parameter, $a$, of the bond fluctuation model is computed from the Lennard-Jones parameter, $\sigma$, for polystyrene~\cite{Rossi2011}. The minimum energy well of the Lennard-Jones potential corresponds to the minimum interaction distance, $2\,a$, of the bond fluctuation model, so that $2\,a={2^{1/6}}\sigma=4.7$ \AA. The distance and energy parameters associated with each pair-potential for the two systems considered (polystyrene in cyclohexane and polystyrene in toluene) are listed in Table~\ref{PairPotentials} and plotted in Figure~\ref{Potentials}. The plotted points in the figure represent the discrete interaction distances allowed under the bond fluctuation model and their corresponding energies. A single value of $\sigma$ was adopted for all the pair potential functions (essentially a coarse-graining assumption) to reduce the bias introduced by the limited number of interaction distances allowed under the bond fluctuation model.

\begin{table*}[htbp]
\caption{Pair-potential functions used in the bond fluctuation model for the polystyrene system.}
\label{PairPotentials} 
 \begin{threeparttable}
    \centering
\footnotesize
\begin{tabularx}{0.99\textwidth} { 
  | >{\raggedleft\arraybackslash}c 
  | >{\raggedright\arraybackslash}c 
  | >{\raggedright\arraybackslash}X 
  | >{\raggedright\arraybackslash}c| }
  \hline
 & $n,m$ & \multicolumn{1}{|c|}{pair} &  \\
 & \multicolumn{1}{|c|}{molecule} & \multicolumn{1}{|c|}{potential} &  \\ 
\multicolumn{1}{|c|}{ $V^{\phi}_{n,m}$ } & \multicolumn{1}{|c|}{pair types} & \multicolumn{1}{|c|}{function} & \multicolumn{1}{|c|}{parameters} \\[2mm] \hline
$V_{\text{pp}}$ & monomer:monomer &\makecell{ $\; 4 \varepsilon _{\alpha\beta}\left(\left(\frac{\sigma }{r_{\text{}}}\right)^{12}-\left(\frac{\sigma }{r_{\text{}}}\right)^6\right)$} & $\sigma=4.2$\AA, 
$\frac{\varepsilon_{\text{pp}} }{ k_{\text{\tiny b}} }= 408K$ \tnote{a} \rule[-.3\baselineskip]{0mm}{9mm} \\[6mm]

$V_{\text{ps}}$ & monomer:solvent &\makecell{ \; $4 \varepsilon _{\alpha\beta}\left(\left(\frac{\sigma }{r_{\text{}}}\right)^{12}-\left(\frac{\sigma }{r_{\text{}}}\right)^6\right)$} &\makecell{ p--cyclohexane (0.5): $\frac{\varepsilon_{\text{ps}}}{k_{\text{\tiny b}}}=380K$\\[2mm]
p--toluene ($-1.25$): $\frac{\varepsilon_{\text{ps}}}{k_{\text{\tiny b}}}=330K $\\[2mm]
p--toluene (1.75): $\frac{\varepsilon_{\text{ps}}}{k_{\text{\tiny b}}}=480K $ } \tnote{b} \rule[-.3\baselineskip]{0mm}{9mm} \\[6mm]

$V_{\text{ss}}$ & solvent:solvent &\makecell{ $\; 4 \varepsilon _{\alpha\beta}\left(\left(\frac{\sigma }{r_{\text{}}}\right)^{12}-\left(\frac{\sigma }{r_{\text{}}}\right)^6\right)$} & \makecell{ cyclohexane: $\frac{\varepsilon_{\text{ss}}}{k_{\text{\tiny b}}}=297K$\\[2mm]
toluene: $\frac{\varepsilon_{\text{ss}}}{k_{\text{\tiny b}}}=377K $ } \tnote{c} \rule[-.3\baselineskip]{0mm}{9mm} \\[6mm]

$V_{\text{FENE}}$ & intrachain monomers &\makecell{$\; -\frac{1}{2} K R_{0}^{2} \ln \left(1-\left(\frac{r}{R_{0}}\right)^2\right)$}
 & $R_0=a\sqrt{11}$ \rule[-.3\baselineskip]{0mm}{9mm} \\ 
 &  &\makecell{$ + 4\, \varepsilon \left(\left(\frac{\sigma }{r}\right)^{12}-\left(\frac{\sigma }{r}\right)^6\right)+\varepsilon$} & $K=45,\frac{\varepsilon}{k_{\text{\tiny b}}}=\frac{\varepsilon_{\text{pp}}}{k_{\text{\tiny b}}}= 408K$
 \tnote{d} \\[6mm] 
$V_{\text{WALL}}$ & monomer:wall &\makecell{ $\; \varepsilon \left(\frac{2}{15} \left(\frac{\sigma }{r_{\text{}}}\right)^9-\left(\frac{\sigma }{r_{\text{}}}\right)^3\right)$} \rule[-.3\baselineskip]{0mm}{9mm}& $\sigma=\frac{2}{2^{1/6}}, \;\;\; \frac{\varepsilon}{k_{\text{\tiny b}}}=\frac{\varepsilon_{\text{pp}}}{k_{\text{\tiny b}}}= 408K$ \rule[-.3\baselineskip]{0mm}{5mm} \\ 
\hline
\end{tabularx}

\begin{tablenotes}
  \item[a] These values are from references~\cite{Jimenez-Serratos2017,Xiao2016}.
  \item[b] These are the calculated heterogeneous interaction parameters from Equation~(\ref{FloryHuggins}); the numbers in parentheses represent the value used for the Flory-Huggins parameter, $\chi$~\cite{Mark2006}.
  \item[c] These values are from references~\cite{Clancy2001,Galliero2005,Poling2001}.
  \item[d] The spring constant, $K$, is increased from the literature value of $30$ to prevent unnatural chain compression~\cite{Dimitrov2007, He2007} and is consistent with bond energy scaling values 3 to 4 orders of magnitude greater than Lennard-Jones interactions~\cite{Rossi2011}.
  \item[e] These values are from reference~\cite{Dimitrov2007}.
\end{tablenotes}
\end{threeparttable}

\end{table*}

The interaction parameters, $\varepsilon_{{\alpha}\beta}$, in Table~\ref{PairPotentials} between monomer and  solvent can be estimated using the Flory-Huggins parameter, $\chi$. The Flory-Huggins solution parameter for the enthalpy of mixing can be related to the interaction parameters among the solvent and polymer species~\cite{Huggins1964, Sariban1987, Mark2006} via
\begin{equation}
\chi=\frac{N_c}{k_{\text{\tiny B}} T} \left(\varepsilon_{\alpha\beta}-\frac{1}{2}\left({\varepsilon_{\alpha\alpha}+\varepsilon_{\beta\beta}}\right)\right)
\label{FloryHuggins}
\end{equation}
where $\varepsilon_{\alpha\beta}$ represents the interaction energy between two different molecular constituents, while $\varepsilon_{\alpha\alpha}$ and $\varepsilon_{\beta\beta}$ represent self-interaction energies. The quantity, $N_c$, is the coordination number for the lattice. The Flory-Huggins parameter was derived for a simple cubic lattice and excludes the bonded monomers present in the interactions. A value of 6 is used for $N_c$~\cite{Miquelard-Garnier2016}.   
Equation~ (\ref{eqn:Ej}) provides the energy of any possible brush conformation. Using the bond fluctuation model with the parameters in Table~\ref{PairPotentials}, the bond energy of each molecule in a configuration is obtained using the five possible pair potentials between a molecule and its neighbors. All the interacting neighbors lie within a distance of $4.7$,  $5.3$,  $5.8$,  $6.6$,  $7.1$, or  $7.4$ \AA~ of the reference molecule. These are the interaction distances allowed by the lattice of the bond fluctuation model, and they constitute the terms of Equation~(\ref{eqn:Ej}) in the parentheses. The energy of the whole brush conformation is obtained by summing the bond energies of all the molecules in the simulation volume (the outer sum in Equation~(\ref{eqn:Ej})).

Another important structural characteristic of polymer brushes is the grafting density, $\sigma_d$. It is defined as the number of chains per unit area of the grafting surface and can be calculated from the simulation domain using 
\begin{equation}
\sigma_d = \frac{N_{\text{chain}}}{L_x \, L_y}
\end{equation}
where $N_{\text{chain}}$ is the number of polymer chains on the grafting plane, and ${L_x}$ and ${L_y}$ are the lateral dimensions of the simulation block. For the present simulation, the lateral dimensions are both 17 lattice parameters so the grafting density is $\sigma_d = 8/(17)^2$ or 3\%. 

For the simulation domain of Figure~\ref{ExampleBrush} with 3\% grafting density and the parameters of Table~\ref{PairPotentials}, Equation~(\ref{eqn:Ej}) can be used to calculate the energy of an arbitrary conformation of the styrene brush in either cyclohexane or toluene solvent. This equation provides the basis for generating a degeneracy of the energy eigenlevels --- a map of the energies for all the possible conformations of the brush + solvent system. Because the system is discretized by the form of Equation~(\ref{eqn:Ej}), the energy of any state of the system can be expressed as the expectation value of the occupied energy levels of this landscape.  

To predict the kinetic evolution of the brush system according to the steepest-entropy-ascent principle, it is imperative to determine the entropy of the system at each state. This can be achieved by directly calculating the entropy from the occupied energy levels and their degeneracies. Fortunately, the Replica Exchange Wang-Landau algorithm offers a computationally efficient way to estimate the degeneracies of all energy levels. 

\subsection{Replica Exchange Wang-Landau} \label{subsec:WangLandau}

\begin{figure}
	\begin{center}
		\includegraphics[width=0.45\textwidth]{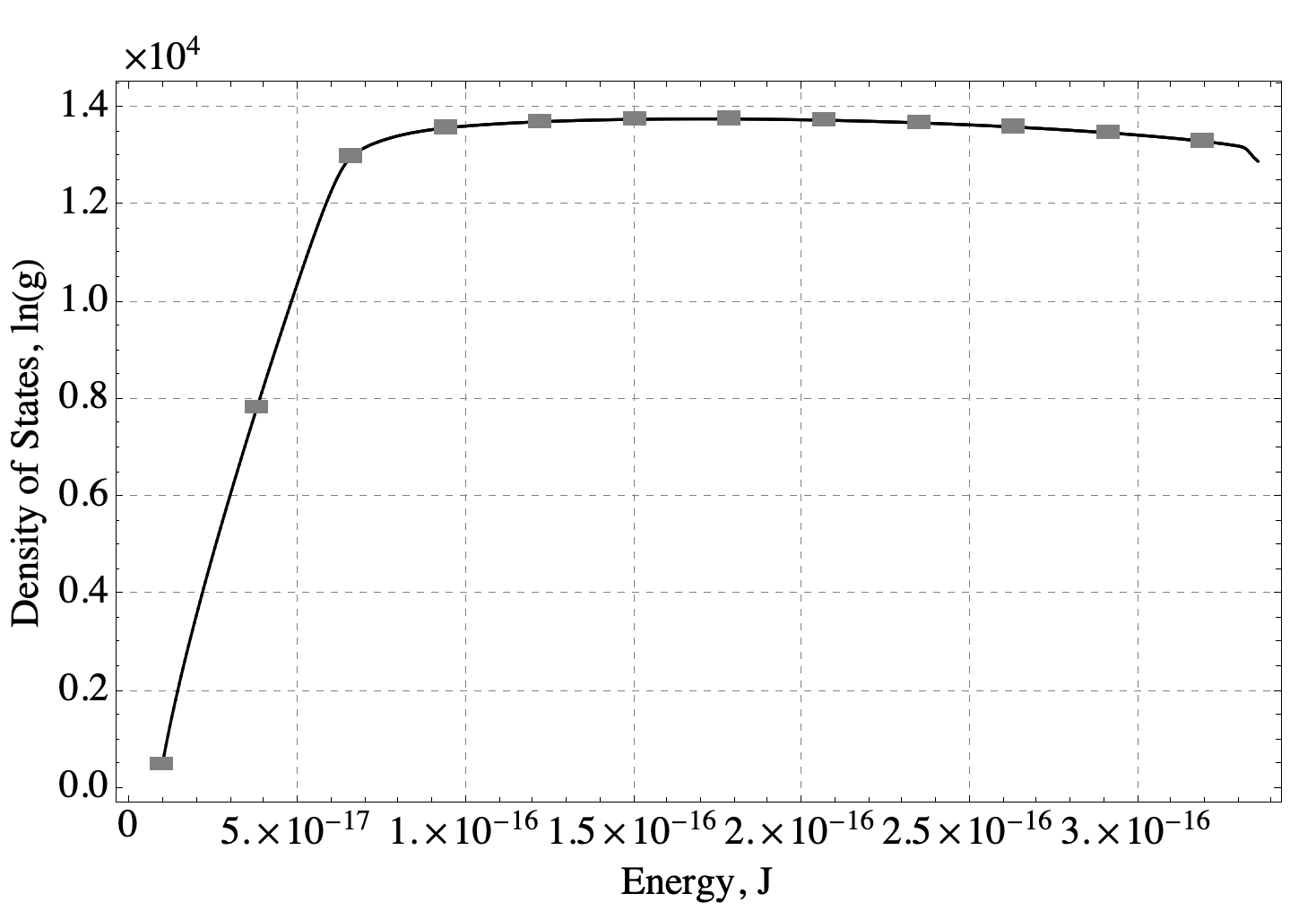}
		\caption{The energy eigenstructure estimated with the Replica Exchange Wang-Landau algorithm for a styrene polymer brush in a cyclohexane solvent like the one shown in Figure~\ref{ExampleBrush}. The discrete energy levels, $E_j$,  for all possible brush conformations are plotted on the horizontal axis, and the natural logarithm of the degeneracy, $g$, corresponding to each energy level is plotted on the vertical axis.} 
		\label{DOS}
	\end{center}
\end{figure}

The energy eigenstructure of a system comprises the discrete spectrum of all possible energy levels and the degeneracy of each energy level (i.e., the number of eigenstates with the same energy). This landscape is also known as the system's density of states. Wang-Landau is an algorithm recently developed for estimating the energy eigenstructure~\cite{WangLandau2001a, WangLandau2001b, Li2014, Vogel2013}. Figure~\ref{DOS} shows an example of such a landscape for a styrene polymer brush in a cyclohexane solvent. The lowest energy is about 0.1 x $10^{-16}$ J (or 51.3 J/mol), while the highest is approximately 3.4 x $10^{-16}$ J (or 1745.5 J/mol). The highest degeneracy in this figure is $\ln(g)\approx 14,000$. The energy level with that degeneracy has a hyper-astronomical number of eigenstates of $g=e^{14,000}\approx 10^{6080}$. While numbers this large cannot be physically realized, they are common when considering degeneracies in systems with many molecules. It is important to estimate these degeneracies accurately in order to reliably calculate the entropy of each energy level. The Wang-Landau algorithm employed to do this is based on a non-Markovian Monte Carlo method that estimates the degeneracies of the energy eigenlevels from a uniform sampling of the system's {\em energy levels} (rather than its eigenstates). Wang-Landau relies on the underlying principle that the probability of visiting a given energy level, $p_{j}^s$, is a function of its degeneracy, ${g_i}$, such that $p_{j}^s \propto \frac{1}{g_i}$. The stochastic movements between levels are governed by a corresponding principle in which the probability of movement is related to the relative sizes of a randomly chosen decimal number, $\mathrm{R}$, and the ratio of the estimated degeneracies of the two levels:
$$p^{s}({\varepsilon}_1 \rightarrow {\varepsilon}_2) = \min\left[\frac{g({\varepsilon}_1)}{g({\varepsilon}_2)}\; ,\;0< \mathrm{R}<1 \right] \;\;\;\;\; .$$ This principle allows Wang-Landau to efficiently explore the occupiable energy levels of a system without quasi-thermal constraints like the metropolis transition probability used in Monte Carlo models~\cite{LandauTsaiExler2004}.

The accuracy of the density of states calculation is represented by the error bars shown in Figure~\ref{DOS}.  These correspond to the relative error determined from six runs of the Replica Exchange Wang-Landau algorithm on a 58--monomer chain in previous work~\cite{McDonald2023polymer}. The error bars for the monomer chain represented $\pm$ one standard deviation of the degeneracy determined from the six independent simulations completed with the same Wang-Landau parameters as those used here for the brush system. Because the degeneracy of the monomer chain system and the brush system are different, the error bars in Figure~\ref{DOS} should be treated as an estimate of the error expected in the density of states curve.

Wang-Landau functions by tracking counters that test for convergence to accurate values of the energy degeneracies. During stochastic transition attempts by the algorithm, two vectors are incremented: one is a histogram of Monte Carlo visits to the energy levels, and the other is an estimate of the degeneracies. For each visit to an energy level, the histogram is incremented by ${1}$ and the corresponding degeneracy is multiplied by the natural logarithm of a modification factor, $f$. The value of $\ln(f)$ starts at a value of $1$ and is halved when the histogram reaches a chosen flatness criterion, i.e., when the minimum histogram value is greater than or equal to the average histogram value multiplied by a flatness criterion --- this indicates the energy levels have been sampled uniformly. At this point, the histogram is reset and the process is repeated. The systematic reduction in the modification factor over multiple sweeps through the energy spectrum is needed to obtain degeneracy values with a desired accuracy~\cite{Wust2012, Vogel2013, Vogel2014, Vogel2018, LandauTsaiExler2004}.  In this work, the flatness criterion is set to ${95\%}$ and the algorithm repeated until $\ln(f) \le 10^{-7}$.

The number of energy levels in the spectrum, however, places a practical limitation on the Wang-Landau algorithm. For example, each energy level in a landscape with $10^6$ levels might be visited more than $10^5$ times to reduce the modification factor once. That leads to a cumbersome number of Monte Carlo steps for convergence ($>10^{12}$). The Replica-Exchange variant of the Wang-Landau algorithm improves upon this limitation by using multiple walkers to sample portions of the energy eigenstructure in parallel~\cite{Vogel2013, Vogel2014, Vogel2014Conf, Li2014}. This approach subdivides the landscape into overlapping energy windows with multiple walkers per window.  The histograms and degeneracies of each window are estimated separately. To match degeneracies from one energy window to the next, windows periodically swap or exchange the conformations and estimated degeneracies of the overlapping levels of adjacent windows. This information exchange yields a uniform degeneracy with no abrupt breaks over the whole energy spectrum.  

The Replica-Exchange Wang-Landau code~\cite{Vogel2018} uses integer indices for the histogram and degeneracies to increase computational efficiency. Consequently, the energies calculated from Equation~(\ref{eqn:Ej}) cannot be used directly; they must to be transformed to integer values. To accomplish this, the pair potentials of Table~\ref{PairPotentials} are normalized by the Lennard-Jones parameter for polystyrene. This scaling makes the magnitude of all the pair potentials less than one, so the normalized values are then multiplied by 10 and rounded to the nearest integer. 

The number of energy levels in the spectrum are reduced further by binning closely-spaced energy levels. The bin width here is set to the energy associated with the energy change of an average conformation transition. During an attempted Monte Carlo transition, the difference in energy indexes (after the aforementioned scaling) is $\approx 10$. Energy levels within each range of 10 are grouped together as a single binned energy level. This binning method reduced the energy spectrum to $52,000$ discrete levels without affecting significantly the final degeneracy or the structural parameters calculated from it.

Now, a lattice as opposed to an off-lattice model is employed here for computational efficiency and past precedent. This is justified on the basis of previous work such as that by Farris $et \; al.$\cite{Farris_Seaton_Landau_2019}, who provide a detailed derivation of the degeneracy of a simple polymer structure, using a conventional lattice model as well as a ``continuum,'' off-lattice one. Only first nearest-neighbor energy interactions are considered. Comparing the results for the two lattice models show some differences in the degeneracy curves although the overall curves are quite similar and share several points of commonality, including the same plotted initial, maximum, and end points.

Another point to be made here is that unlike other statistical approaches (e.g., multicanonical), the Wang-Landau algorithm does not require the detailed balance condition for convergence to the density of states~\cite{Wust2012,zhou2008}. It is, however, important that the trial moves used in the algorithm satisfy the detailed balance so as to avoid systematic errors. This is satisfied by using a trial-and-error procedure that chooses trial moves randomly and independently of the current conformation; that is, the algorithm chooses with constant probability \cite{Wust2012}.  Thus, Wang-Landau satisfies the detailed balance asymptotically in the final iterations. In addition, the sampling of the Replica Exchange Wang-Landau and Wang-Landau algorithms has been very carefully tested and the error quantified, showing that the statistical error is generally more problematic than any systematic error relative to convergence to the density of states \cite{Farris2019,Vogel2014,Vogel2013,brown2011}. It is this statistical error, which is quantified below. Furthermore, the Wang-Landau algorithm is only used here to predict the system energy spectrum and degeneracies needed by our equation of motion (discussed in the next section). It is the equation of motion that predicts the kinetics of each non-equilibrium path based on the principle of steepest entropy ascent, not the Wang-Landau algorithm. In addition, for the non-equilibrium states predicted by the equation of motion, not only is a detailed balance not required, it is questionable whether or not such a balance is even valid at non-equilibrium. We, of course, acknowledge that the Wang-Landau and Replica Exchange Wang-Landau algorithms, and indeed all similar statistical methods, are approximate and that there is going to be uncertainty in the density of states generated by them, particularly at the low energy levels. Nonetheless, we note that Farris and Landau \cite{Farris2021replica} have demonstrated that Replica Exchange Wang-Landau compares favorably with the best multicanonical method (PERM in their paper) and the best replica-exchange Monte Carlo method even at the low energy levels.  Nonetheless, granting the unavoidable uncertainties introduced by Replica Exchange Wang-Landau and all other statistical methods particularly at low energy levels for large systems, they do not significantly affect the results presented below because the low energy levels contribute very little to the thermodynamic states associated with the predicted irreversible thermodynamic paths. Thus, our choice of the Replica Exchange Wang-Landau algorithm for the problem at hand is based on a balance between accuracy and computational efficiency~\cite{Hayashi2019,Farris2021replica}. 

A final point is that use of the Replica Exchange Wang-Landau algorithm is not central to our methodology. Other less efficient stochastic algorithms such as the ones mentioned above could be used. Alternatively, a hybrid algorithm such as the multicanonical replica-exchange method constructed by Hayashi and Okamoto \cite{Hayashi2019} could be employed. These authors compared an exact density of states for a square lattice Ising model with simulations from their hybrid algorithm and results from the Replica Exchange Wang-Landau and multicanonical algorithms. They found only a marginal improvement with the hybrid results over those produced by the Replica Exchange Wang-Landau algorithm, while the improvement over the multicanonical results was much more significant. However, the marginal improvement seen with the hybrid algorithm may not justify the loss in efficiency, particularly for the present application since our non-equilibrium results at this point are only qualitative given the lack of available experimental data with which to verify the results.

\subsection{SEAQT Equation of Motion} \label{subsec:SEAQT-EOM}

The SEAQT equation of motion is used to simulate the state-based evolution of a system once the energy eigenstructure has been established. The SEAQT framework is distinct from other methods that attempt to describe kinetic processes, as it derives the kinetics of material systems in a unique way. In Monte Carlo simulations, kinetic evolution is determined by stochastic movements and the gradual decrease in system energy. Constitutive relations or a particle collision model are typically assumed, along with thermodynamic equilibrium information. Molecular dynamics relies on detailed energy descriptions to move hard particles throughout a simulated lattice and generate energy-minimized structures. It relies on a classical mechanics description of particle motion. Although density functional theory (DFT) uses accurate energetic models, it is not designed to follow the kinetic evolution of a system and is more appropriately classified as an optimization procedure for determining the ground-state energy of a configuration.

Steepest-entropy-ascent quantum thermodynamics is a method that is distinct from other approaches in that it has a rigorous thermodynamic basis and utilizes an energy eigenstructure, which is a path-independent entity, to determine a unique kinetic path using its equation of motion. Unlike other methods that require {\em a priori} kinetic, collision, or constitutive models, the SEAQT equation of motion relies on the general principle of steepest entropy ascent at each instant of time, while ensuring that all conservation constraints are satisfied. Additionally, the SEAQT equation of motion is capable of avoiding kinetic trapping, which can occur when a system encounters metastable states. As a result, this equation can generate a unique kinetic path from any initial state that can be described thermodynamically, whether it is in a non-equilibrium or a stable equilibrium state, and guide it through state space to a final state that is either a stable equilibrium state or a steady state.

The equation of motion was originally postulated within a quantum thermodynamic framework, i.e., a unified theory of quantum mechanics and thermodynamics \cite{Hatsopoulos1976-I,Hatsopoulos1976-IIa,Hatsopoulos1976-IIb,Hatsopoulos1976-III, Beretta1984,Beretta1985}, and thus, retains  the matrix formulation of quantum mechanics as well as its representation of states (in this case thermodynamic states) in terms of a set of occupation probabilities of a system’s discrete energy spectrum regardless of whether or not it is applied to a quantum or classical system.  Our application to a polymer brush system is, of course, purely classical and, thus, no quantum effects are taken into account. Furthermore, we note that the SEAQT equation of motion is more general than the von~Neumann equation of quantum mechanics, which is a special case of the SEAQT equation of motion. The dynamics of the von~Neumann equation are strictly linear and limited to the evolution of pure states (zero-entropy states), or mixed states (non-zero-entropy states) for a particular subset of reversible processes, while the dynamics of the SEAQT equation of motion are nonlinear and able to predict the evolution of mixed states for all irreversible processes.  As to the Liouville-Hamiltonian equation, it, like the von~Neumann equation, is strictly reversible and even the transport equations resulting from its irreversible counterpart, the integro-differential Boltzmann equation, have been shown in the low-field limit to be a special case of the SEAQT equation of motion~\cite{Li2018steepest}. In addition, as demonstrated by Beretta~\cite{Beretta2014steepest}, all existing classical non-equilibrium thermodynamic frameworks can be fit into the more general SEAQT framework.

SEAQT has seen many recent advancements in its application to various material systems, including those involving magnetism and spin relaxation~\cite{Yamada2019spin}, thermal expansion~\cite{Yamada2018method}, clustering~\cite{Yamada2020kineticpartII}, nucleation~\cite{Yamada2019kineticpartI}, material coupled electron-phonon flows~\cite{Li2018steepest}, GaN film growth~\cite{kusaba2019,kusaba2017}, electroporation of cell membrane lipid structures~\cite{Goswami2021}, capillary systems~\cite{McDonald2022capillary}, 2D materials~\cite{Younis2022}, and polymer folding~\cite{McDonald2023polymer}. SEAQT is unique in that it can predict the non-equilibrium thermodynamic path that a system takes without requiring knowledge of underlying kinetic processes. This is achieved by assuming that the system's kinetic path follows the principle of steepest entropy ascent, i.e., the path that maximizes entropy production at every instant of time ~\cite{Li2016a,Li2016b,Li2018,Li2018steepest,Beretta2014steepest,Beretta2006,Beretta1984}. The SEAQT equation of motion for a simple quantum system is expressed in operator form as:
\begin{equation}
\frac {d \hat{\rho}} {dt}=\frac {1} {i \hbar}[\hat{\rho},\hat{\mathcal{H}}]+\frac {1} {\tau(\hat{\rho})} {\hat D(\hat{\rho})} \;\;\;\;\; . \label{EOM1}
\end{equation}
In this matrix equation, $t$ is the time and $\hat{\rho}$ is the density or so-called ``state'' operator, which for the classical system modeled here is interpreted as a probability distribution. The first term to the right of the equal is the symplectic term that corresponds to the von~Neumann term of the time-dependent part of the Schr\"{o}dinger equation of motion of quantum mechanics where $[\cdot]$ represents the commutator defined by $[\hat{\rho},\hat{\mathcal{H}}]=\hat{\rho} \, \hat{\mathcal{H}} -  \hat{\mathcal{H}} \, \hat{\rho}$, $\hat{\mathcal{H}}$ is the Hamiltonian operator, and the factors $i$ and $\hbar$ are the imaginary unit and modified Planck constant, respectively. This symplectic term is used in quantum mechanics to predict the reversible evolution of pure states (i.e., zero-entropy states) and is needed in particular when quantum correlations are involved (e.g.,~\cite{Cano2015steepest,jhon2020,jhon2022}). The second term to the right of the equal sign is the dissipation term, which captures system evolution involving the nonzero-entropy states of irreversible processes~\cite{Beretta2014steepest, Beretta2006, Beretta2009, Li2016a}. Here, $\tau$ is a relaxation parameter and $\hat{D}$ the dissipation operator.

For the classical system considered here, both $\hat{\rho}$ and $\hat{\mathcal{H}}$ are diagonal in the energy eigenvalue basis and, thus, commute causing the symplectic term to vanish. The dissipation term $\hat{D}$ can be derived using a variational principle based on a gradient descent in Hilbert space in the direction of steepest entropy ascent~\cite{Beretta2014steepest,Beretta1984}. The energy and occupation probabilities are automatically conserved by ensuring that the equation of motion moves in the steepest-entropy-ascent direction, which is the direction of the component of the gradient of the entropy operator perpendicular to the manifold containing the generators of the motion, i.e., the identity and Hamiltonian operators~\cite{Beretta2014steepest}. For an isolated system, the equation of motion can be expressed as:
\begin{equation}
\frac{dp_j}{dt}=\frac {1} {\tau}\frac{\left|
\begin{array}{ccc}
 -p_j \ln \frac{p_j}{g_j} & p_j & {E}_j\, p_j \\
 \langle S \,\rangle & 1 & \langle E \,\rangle \\
 \langle E\,S \,\rangle & \langle E \,\rangle & \langle E^2 \,\rangle \\
\end{array}
\right|}{\left|
\begin{array}{cc}
 1 & \langle E \,\rangle \\
  \langle E \,\rangle & \langle E^2 \,\rangle \\
\end{array}
\right|} \;\;\;\;\;\; .
\label{EOM2}
\end{equation}
In this expression, the change in the probability of a given energy eigenlevel, $p_j$, in the time interval $dt$ is proportional to a ratio of two determinants. These determinants are functions of the energy, ${E_j}$, and the entropy, $S_j = -\ln \frac{p_j}{g_j}$, of the $j^{th}$ eigenlevel~\cite{Gyftopoulos1997}. They are also functions of the expectation values of the thermodynamic properties $\langle \, \cdot \, \rangle$ obtained from the following definitions:

\begin{align}
\langle E\, \rangle =&\; \underset {j} {{\sum}}\phantom{l} p_j \, E_{j} \nonumber  \\
\langle S\, \rangle =&\; \underset {j} {{\sum}}\phantom{l} p_j \, S_{j} = \underset {j} {{\sum}}\phantom{l} - p_j \, \ln \frac{p_j}{g_j} \nonumber \\
\langle E^2\, \rangle =&\; \underset {j} {{\sum}}\phantom{l} p_j \, E_{j}^{2} \nonumber \\
\langle E\,S\, \rangle =&\; \underset {j} {{\sum}}\phantom{l} - p_j \, E_j \ln \frac{p_j}{g_j} \;\;\;\;\;  .  
\label{ExpectationValues}
\end{align}

Now, although Equation~(\ref{EOM2}) is limited to an isolated system, such a system can be partitioned into any number of open subsystems that can account for the transfer of, for example, heat and mass between subsystems within the overall isolated system. To do this, we assume each non-equilibrium state can be described by a 2$^{nd}$-order hypo-equilibrium state~\cite{Li2016a}. In general, an $M^{th}$-order hypo-equilibrium state results from a subdivision of the state space (e.g., Hilbert space) of a system into $M$ subspaces, which may or may not coincide to physical subsystems. For example, there is a direct correspondence between subspace and subsystem in the case of the transfer of energy or mass within the system but no such correspondence in the presence of a chemical reaction. Furthermore, the occupation probabilities of each subspace at any given instant of time are described by a canonical distribution and the combination of canonical distributions of all subspaces at a given instant of time represents the non-equilibrium state of the system. Thus, hypo-equilibrium states are a subset of non-equilibrium states.  In addition, it has been proven that they can approximate to a high degree of accuracy any non-equilibrium state, and that the order of the hypo-equilibrium state is preserved by the SEAQT equation of motion throughout its time evolution~\cite{Li2016a}. Once each non-equilibrium state is described by a 2$^{nd}$-order hypo-equilibrium state, the isolated system can be subdivided into subsystems $A$ and $B$. The relevant equation of motion for subsystem ${A}$ for the case when only energy is exchanged between subsystems $A$ and $B$ becomes
\begin{equation}
\frac{dp^A_j}{dt}=\frac {1} {\tau}\frac{\left|
\begin{array}{cccc}
 -p_j^A \text{ln}\frac{p_j^A}{g_j^A} & p_j^A &0 & {E}_j^A p_j^A \\
 \langle S^A \,\rangle & 1 & 0 &\langle E^A \,\rangle \\
 \langle S^B \,\rangle & 0 & 1 &\langle E^B \,\rangle \\
 \langle E\,S \,\rangle & \langle E^A \,\rangle & \langle E^B \,\rangle & \langle E^2 \,\rangle \\
\end{array}
\right|}{\left|
\begin{array}{ccc}
 1 & 0&
\langle E^A \,\rangle \\
  0&1&\langle E^B \,\rangle \\
   \langle E^A \,\rangle &\langle E^B \,\rangle &\langle E^2 \,\rangle\\
\end{array}
\right|}  \;\;\;\;\;\; .
\label{ABEqM}
\end{equation}
The determinant in the numerator can be re-written as a row expansion with the co-factors of the first row denoted by $C_1$, $C^A_2$, $C_3$. The equation of motion for subsystem $A$ can then be expressed in the equivalent form as
\begin{align}
\frac {dp_j^A}{dt^*} & = 
p_j^A \left(-\text{ln}\frac{p_j^A}{g_j^A}-\frac{C_2^A}{C_1}-{E}_j^A \, \frac{C_3}{C_1}\right) \nonumber \\
&   =  
p_j\left[\left(S_j^A - \langle S^A \,\rangle \right)-\left({E}_j^A - \langle {E}^A\rangle \right)\frac{C_3}{C_1}\right] 
\label{AEqMCoF}
\end{align}
with $t$ and $\tau$ replaced by a dimensionless time defined as $t^*= \int_0^t \frac{1}{\tau(\vec{p}(t'))}dt'$. The relaxation parameter, $\tau$, can either be assumed constant or taken to be a function of the time-dependent occupation probabilities $p_j$ represented by the vector $\vec{p}$. If it now is assumed that the size (energy) of subsystem $B$ is much greater than that of $A$ and that as a result $B$ remains in a stable equilibrium state, subsystem $B$ can act as a heat source or sink indistinguishable from a thermal reservoir~\cite{Li2016a,Li2016b,Li2018}. In this case, the co-factor term remaining in Equation~(\ref{AEqMCoF}) can be replaced with the reservoir temperature, $T^R$, such that
\begin{align}
\beta^R = \frac{C_3}{C_1}= \frac{1}{k_b T^B}= \frac{1}{k_b T^R}
\end{align}
The equation of motion for subsystem $B$ is  no longer needed because subsystem $B$ remains in a stable equilibrium state.

Equation~(\ref{AEqMCoF}) defines a set of first-order, ordinary differential equations that describe the dynamics of the system over time. Its solution provides the occupation probability for each possible energy eigenlevel at each instant of time~\cite{Li2016a, Li2016b}. For the polymer brush and solvent system considered here, the system is described by approximately 52,000 equations of motion and an equal number of unknowns. This system of equations is solved using Matlab's ODE45 numerical solver.

To utilize the equation of motion, Equation~(\ref{AEqMCoF}), an initial state must be specified. In this context, a thermodynamic state is specified by a probability distribution for the occupiable energy eigenlevels of the system. One strategy for arriving at such a state, which we employ here, is to initially generate a partially canonical state (e.g., a metastable or unstable equilibrium state) and subsequently perturb it to produce an initial non-equilibrium state. To achieve this, we use the following equations to describe the occupation probabilities of a canonical (stable equilibrium), $p_j^{se}$, and a partially canonical (partial equilibrium), $p_j^{pe}$, state:
\begin{equation}
p_j^{\text{se}}=\frac {g_j \exp(-\beta^{\text{se}} E_j)}{\underset {j} {{\sum}}\phantom{l}g_j \exp(-\beta^{\text{se}} E_j)}
\label{CanDist}
\end{equation}
\begin{equation}
p_j^{\text{pe}}=\frac {{\delta}_j \, g_j \exp(-\beta^{\text{pe}} E_j)}{\underset {j} {{\sum}}\phantom{l}{\delta}_j \, g_j \exp(-\beta^{\text{pe}} E_j) } \;\;\;\;\;\;\;  .
\label{PartCanDist}
\end{equation}
In the preceding equations, $g_j$ denotes the degeneracy associated with the $j^{th}$ energy eigenlevel, while $\beta^{\text{se}}$ is inversely proportional to the temperature of the canonical state and $\beta^{\text{pe}}$ is simply a parameter of the partially canonical state that needs to be determined.   Additionally, the variables $\delta_j$ take on values of either 0 or 1, with different sets of $\delta_j$ values corresponding to different partially canonical states. To obtain the canonical and partially canonical probabilities for a fixed system energy, $\langle E \rangle$ (obtained from Equation~(\ref{ExpectationValues})) and fixed canonical temperature (i.e., a fixed $\beta^{\text{se}}$), Equations~(\ref{CanDist}) and (\ref{PartCanDist}) are solved for the unknown probabilities and $\beta^{\text{pe}}$. Once determined, the perturbation function $p_j = p_j^{{pe}}\lambda + p_j^{{se}}(1-\lambda)$ is used to generate an initial non-equilibrium probability distribution. Here, $\lambda$ is a number between 0 and 1.

Note that this procedure is used because it is systematic and provides either an initial non-equilibrium state or a metastable equilibrium state. The latter can be used as an initial state when the system interacts with a reservoir. This procedure allows one to more easily control the location of the initial state within the non-equilibrium region  (e.g., far from equilibrium as opposed to near to equilibrium). Of course, any non-equilibrium or metastable equilibrium state can be used regardless of whether or not it was produced by this procedure. The equation of motion is independent of the procedure and is extremely robust and will converge to a stable equilibrium state regardless of the initial state used provided that that state is not a pure state (a zero-entropy state). In that case, the equation of motion reduces to the von~Neumann equation of motion and the symplectic term of the SEAQT equation of motion appears while the dissipation term vanishes.

\subsection{Linking State Space to Microstructural Evolution} \label{sec:MicroLink}

While the SEAQT equation of motion is computationally efficient for predicting the evolution of a system without real-space conformational changes, it would be beneficial to establish a connection between the predicted thermodynamic states and physical conformations in a post-processing step. Stochastic methods like Monte Carlo simulations can provide such a connection, but they are computationally expensive and do not account for non-equilibrium paths that a system can take. Additionally, these methods can trap a system in a metastable well, leading to incorrect information about relaxation times and the system's evolution. The SEAQT framework avoids these issues by adhering to the Lyapunov stability requirement outlined in the Hatsopoulos-Keenan statement of the second law of thermodynamics~\cite{Hatsopoulos1965,Gyftopoulos2005}, which states that the maximum entropy states are the only equilibrium states of the dynamics that are unconditionally stable as defined in~\cite{Beretta1986lyapunov}. In other words, these states are globally stable whereas the metastable states are only locally so. Thus, SEAQT dynamics move the system from some non-equilibrium state to stable equilibrium.

Recent research has established a link between SEAQT paths in state space and physical conformations~\cite{McDonald2022capillary, McDonald2023polymer}. To accomplish this, representative microstructures or conformations are connected to SEAQT-predicted state evolution through the expectation values of a set of structural parameters or descriptors that characterize the microstructure at each time step. These expectation values are computed using the predicted probability distribution and the arithmetic average values of these parameters determined for each eigenlevel during the execution of the Wang-Landau algorithm. In the present study, the two structural parameters used are the radius of gyration, $R_g^2$, and the tortuosity, $\zeta$.

The radius of gyration is a commonly used characterization parameter for simulated polymer systems, as it provides a quantitative measure of the relative clustering and change in position of the center of mass. It can be experimentally determined via scattering techniques. In this work, the radius of gyration is calculated separately for each chain (all chains have the same mass) using the equation
\begin{equation}
R_g^2 = {\left(\frac{1}{N} \sum_i^N   \lVert \boldsymbol{r}_i-\boldsymbol{r}_{\textrm{cm}}\rVert{}^2\right)}
\label{Rg}
\end{equation}
where $N$ is the number of monomers in the chain, $r_i$ represents the 3D vector coordinate of a given monomer, and $r_{cm}$ is the position of the center of mass of the chain. The calculated radii of gyration for each chain are then averaged to produce a single value.

The tortuosity describes the complexity of the brush conformation. It is defined as
\begin{equation}
\zeta ={\left(\frac{1}{N-2} \sum _i^{N-2} \lVert\boldsymbol{s}_i-\bar{\boldsymbol{s}}\rVert{}^2\right)}^{\frac{1}{2}}
\end{equation}
where $\boldsymbol{s}_i$ is defined as
\begin{equation}
\boldsymbol{s}_i= \left( {\begin{array}{cc}
   {\sum_{j_{\boldsymbol{x}}=1}^i w_{j_{\boldsymbol{x}}}} \\
   {\sum_{j_{\boldsymbol{y}}=1}^i w_{j_{\boldsymbol{y}}}} \\
   {\sum_{j_{\boldsymbol{z}}=1}^i w_{j_{\boldsymbol{z}}}} \\
  \end{array} } \right)
  \end{equation}
and
  \begin{equation}
\boldsymbol{w}_j=\boldsymbol{r}_{j,j+1}\times \boldsymbol{r}_{j,j+2} , \; 1 \leq i \leq (N-2) \;\;\;\;\; .
\end{equation}
To calculate $\boldsymbol{s}_i$, first, the three components of $\boldsymbol{w}$ are determined by taking the cross product of the coordinate vectors of a given monomer and the two succeeding monomers in a sequence. Then, the ${w_x}$, ${w_y}$, and ${w_z}$ components are summed to obtain $\boldsymbol{s}_i$. The tortuosity is computed as the sum of vectors normal to the planes defined by successive turns in the chain and is a measure of the degree of equivalent turns along short segments of the chain, which typically increases during coiling. It is worth noting that if no turns occur across three monomers, the cross product for $\boldsymbol{s}_i$ will equal $(0,0,0)$.

The process of assigning a representative conformation to a thermodynamic state can be broken down into several steps. Firstly, for each conformation explored by the Replica-Exchange Wang-Landau algorithm during the generation of the energy eigenstructure, the radius of gyration and tortuosity are calculated, and the arithmetic average values of these parameters are determined for each energy level. Secondly, the SEAQT equation of motion is executed to identify the energy levels that are occupied along the kinetic path; this is a small subset of all possible energy levels in the system. Thirdly, the Replica-Exchange Wang-Landau algorithm is run again, and conformations associated with Monte Carlo visits to the small subset of occupied energy levels are stored. Lastly, representative conformations along the kinetic path are obtained by selecting conformations from the stored set that have $R_g^2$ and $\zeta$ values that match the expected values of these parameters. This produces a set of representative conformations that smoothly vary along the kinetic path in a manner consistent with the physical descriptors used to characterize the system.

In addition to radius of gyration and tortuosity, other structural parameters can be used to describe morphological changes in brushes. Density profiles, which express the fraction or percentage of a constituent (solvent or polymer) at each $z$-plane from the grafting wall, can be measured from scattering experiments or calculated from computational studies. Simulation data can also be analyzed to extract parameters such as brush height and width. Each of these parameters can be linked with kinetic paths in state space using the probabilities predicted by the SEAQT equation of motion at each instant of time and the arithmetic-averaged values developed by the Replica-Exchange Wang-Landau algorithm for each energy eigenlevel of the energy eigenstructure~\cite{Dimitrov2007,vanEck2020}.

The planar density is a function of the distance, $z$, from the grafting plane and is given by:
\begin{equation}
\phi_{i}(z) = \frac{N_i}{L_x \, L_y} 
\end{equation}
where $\phi_{i}(z)$ is the number, $N_i$, of molecules of the $i$ specie ($i$ is either polystyrene, $p$, or solvent, $s$) on the $z$-plane of the brush divided by the planar area. 

The equation for the brush height, $H$, is defined by the first moment of the density profile curve and is given by~\cite{Dimitrov2007,Karim1994}
\begin{equation}
    H = \frac{\int_{0}^{L_z} z \, \phi_p(z) \,dz}{\int_{0}^{L_z} \phi_p{(z)} \,dz}
\end{equation}
where $L_z$ is the $z$ component of the linear dimensions of the system and $\phi_p(z)$ is the density profile of the brush as a function of depth $z$. 
The brush height describes the extension of the chains in the brush from the grafting surface.

The interfacial width, $W$, is defined by the maximum value of the density profile of the polymer divided by the maximum absolute value of the tangent of the profile corresponding to the inflection point. It is expressed~\cite{vanEck2020} as
\begin{equation}
     W = \frac{ \phi_p{^{\text{max}}}{(z)} }{ \max|\frac{d\,\phi_p{(z)}}{{dz}}|} \;\;\;\;\; .
\end{equation}
For the density calculations, a moving average over two $z-$planes is used to smooth-out fluctuations on alternating planes associated with the occupation constraints of the bond fluctuation model.

\subsection{Notes on Temperature and the Relaxation Parameter \label{Sec:TempTau}}

Along the kinetic path predicted by the SEAQT equation of motion, each state, except the final state, is a non-equilibrium state for which temperature is not defined. However, the final state is a canonical stable equilibrium state where temperature exists. Temperature is thermodynamically defined by $\left(\partial E/\partial S\right)_{V,\textbf{\textit{n}}}$, where $E$, $S$, $V$, and $\textbf{\textit{n}}$ are the expectation values of the system energy, entropy, volume, and composition, respectively. The temperature value depends on the energy and entropy values, as well as the energy model or landscape used. In the current energy eigenstructure, the vibrational and translational energies are excluded from the energy calculation, which affects the reference values for entropy through their influence on the density of states. As a result, the reported temperature values in this work are incomplete and not directly comparable to experimental measurements. Nonetheless, these values provide relative evidence for conformal changes of the material system with temperature.

Regarding the relaxation parameter $\tau$, its value(s) dictate(s) the speed at which the system evolves along the unique non-equilibrium path predicted by the SEAQT equation of motion. As this evolution occurs through state space and not the phase space utilized in molecular dynamics and Monte Carlo simulations, the relaxation parameter must be tied to either experiment or a theoretical yet semi-empirical model that correlates changes in state to specific constituent movements between precise locations within a representative conformation. One such method involves using proportionalities from Rouse dynamics that relate changes in the component of the radius of gyration perpendicular to the grafting plane to the relaxation parameter~\cite{Binder1994} (an analogous method was used to determine $\tau$ for a free polymer chain~\cite{McDonald2023polymer}). Specifically, the relationship for the relaxation parameter can be expressed as
\begin{equation}
\tau = \frac{R_{g_\perp}^2}{D}
\label{tau}
\end{equation}
where
$$
R_{g_\perp}^2 = 
\left(\frac{1}{N} \sum_i^N (\lVert \boldsymbol{r}_{i_x}-\boldsymbol{r}_{\textrm{cm}_x}\rVert^2 + \lVert \boldsymbol{r}_{i_y}-\boldsymbol{r}_{\textrm{cm}_y}\rVert^2)\right) 
$$
Note that $R_{g_\perp}^2$ is dependent only on the the ${x}$ and ${y}$ components of the ${r_i}$ vectors and ${r_{cm}}$. The $D$ in Equation~(\ref{tau}) is an effective diffusion constant of the center of mass of the individual chains and can be related to the solvent viscosity $\eta_s$ such that $D\propto(\eta_s R_g)^{-1}$. The diffusion constant also varies with chain length as $D\propto(\frac{1}{N})$ or it can be taken from literature values. Further details of the choice of the $\tau$ used here are reported in Section~\nameref{section:Discussion}~\cite{Binder1994, Gulari1979, Rossi2011}.

Another point is that when there are multiple frequencies contributing to the time relaxation parameter in the SEAQT equation of motion, a couple of different approaches can be used to account for the effect of these frequencies. The simplest is to use Matthiessen’s rule \cite{chen2005}, which is used to combine the effects of the frequencies on $\tau$. In fact, in polymer brush systems, even single grafted ones \cite{Koch1997}, there is a second time constant parallel to the substrate. However, we expect the perpendicular component of the time constant to be much smaller than that of the parallel one in which case the perpendicular contribution will dominate since $1/\tau = 1/\tau_\perp + 1/\tau_\parallel$ \cite{Binder2011}. The more involved approach is to determine a different relaxation parameter for each energy eignelevel as is done, for example, by one of the co-authors \cite{Li2018,Wardon2023}. In this case, $\tau_\perp$ and $\tau_\parallel$ could be individually assigned to different ranges of the energy eigenspectrum based on the characteristic energy associated with each relaxation parameter. The relaxation parameter $\tau_j = \tau(E_j)$ of each energy eigenlevel would then be found using either $\tau_\perp$ or $\tau_\parallel$ and an appropriate characteristic length and mass and the eigenenergy $E_j$. 

Now, due to the qualitative nature of the results presented here, the simplest approach of employing a constant value of $\tau = \tau_\perp$ as given by Equation~(\ref{tau}) is used. However, regardless of the approach utilized, the kinetic predictions of the equation of motion do not depend on the value of $\tau$. This equation predicts a unique kinetic path independently of $\tau$. The relaxation parameter is simply used in a post-processing step to determine how fast the system proceeds along the predicted path.

A final note is that the scaling provided by the relaxation parameter,  $\tau$, is a common feature of dynamic models at all levels of description. At a continuum level, balances for mass, momentum, and energy require experimental values of the transport coefficients, which necessarily scale the predictions made to real time. This is also true at the mesoscopic level, in which the relaxation parameters appearing in the Boltzmann transport equations, the Bhatnagar–Gross–Krook equation, and the Fokker-Planck equation (via its drift and diffusion coefficients) require either experimental or semi-empirical values for the relaxation parameters to place the predictions made in real time. In fact, as is shown in reference \cite{Li2018steepest}\mbox{}, the Boltzmann transport equations are simply a special case of the SEAQT equation of motion. One of the co-authors has just recently shown, for example, that using experimentally determined characteristic electron and phonon relaxation times specific to a given thermoelectric material, the thermal and electrical properties of a variety of semiconductor materials across a wide range of temperatures can quite accurately be predicted by the SEAQT equation of motion \cite{Wardon2023}. The relaxation parameters, as already noted, do not predict the trends of these properties but instead simply scale them to real time. At a quantum level, this is also true of the Lindblad equation, which, unlike the Schr\"{o}dinger equation of motion, also requires experimental values for its relaxation parameters. The reason that a reversible equation such as the Schr\"{o}dinger equation of motion has no relaxation parameter is that the predictions made are not in real-time since it predicts the motion of a reversible process. In fact, the Schr\"{o}dinger equation is unable to correctly predict the behavior of a quantum computer due to the presence of dephasing and relaxation, which, however, both the Lindblad and SEAQT equations can~\cite{jhon2020,Cano2015steepest,jhon2022}, but that requires values for their relaxation parameters scaled to the experimentally determined characteristic times for dephasing and relaxation specific to a given quantum computer. Of course, it is this relaxation parameter that makes the dissipation term of our equation of motion phenomenological. The dissipation operator itself, on the other hand, is not since it is based on a quantum thermodynamic non-linear dynamics in state space specific to the steepest-entropy-ascent principle that fundamentally captures the quantum or classical dynamics of a given system.

\section{Results} \label{sec:results}
\subsection{Cyclohexane Solvent} \label{subsec:CyclohexaneSolvent}
\subsubsection{Evolution of the System Energy, Entropy, and Entropy Generation}

\begin{figure}
	\begin{center}
	    \includegraphics[width=0.45\textwidth]{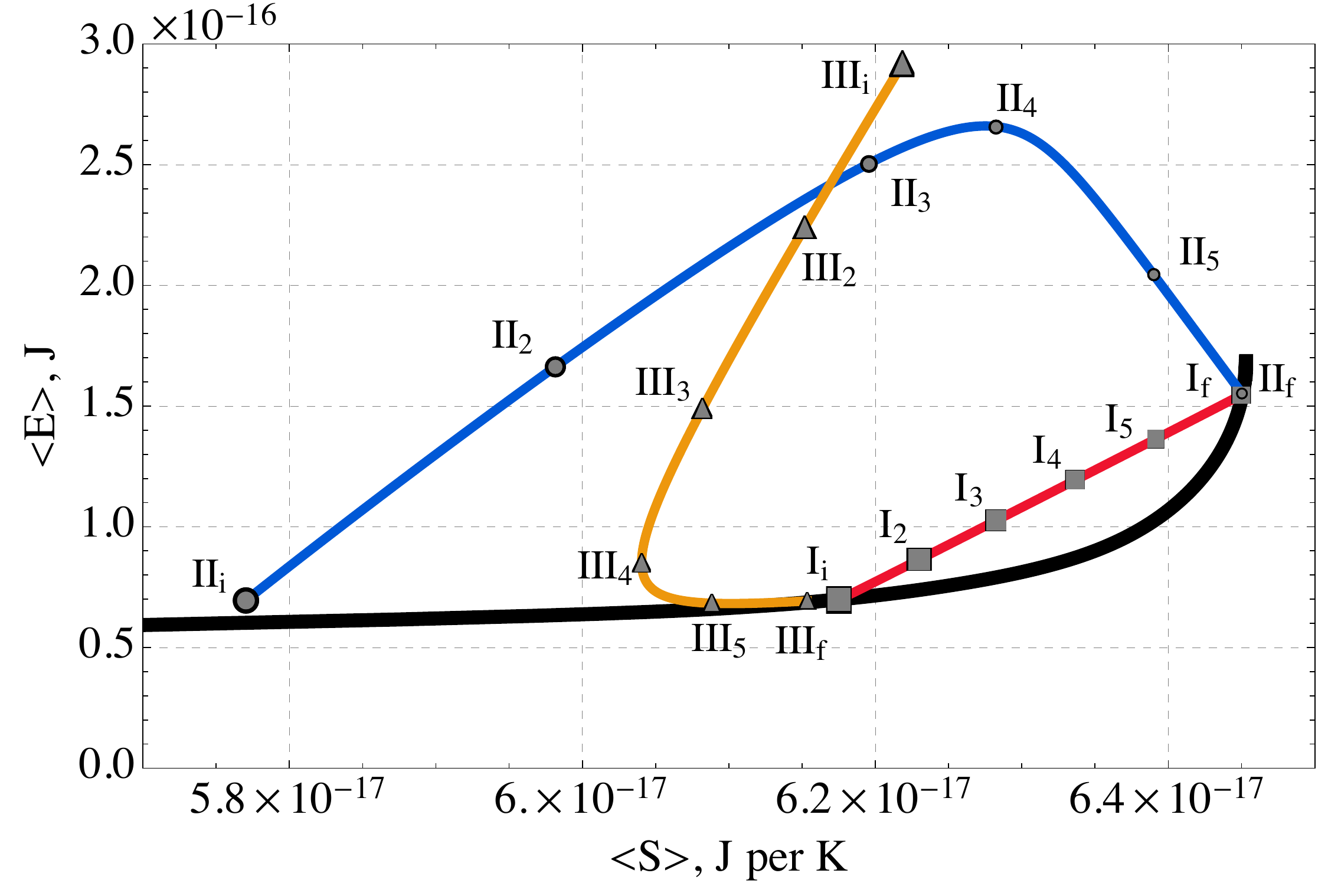}
		\caption{Non-equilibrium thermodynamic paths for the polystyrene-cyclohexane system. The red curve (labeled I) represents a heating path from an initial stable equilibrium state at 6 degrees (arbitrary temperature scale) to a final state in mutual equilibrium with a thermal reservoir at 300 degrees. The blue curve (labeled II) is another non-equilibrium heating path beginning from a low energy non-equilibrium state (close to equilibrium) to a final state in mutual equilibrium with a thermal reservoir at 300 degrees. The orange curve (labeled III) represents a non-equilibrium cooling path starting from a non-equilibrium state going to a final state in mutual equilibrium with a thermal reservoir at 5 degrees.}
		\label{Cyclo_Paths}
	\end{center}
\end{figure}

Energy and entropy are useful parameters to specify equilibria and track kinetic processes of a system. Figure~\ref{Cyclo_Paths} illustrates a map of energy and entropy for the polystyrene brush in cyclohexane system. The horizontal axis shows the expectation value of the entropy, and the vertical axis represents the expectation value of the energy calculated from energy eigenlevels of Equation~(\ref{eqn:Ej}) and a set of occupation probabilities through Equation~(\ref{ExpectationValues}). The black curve represents the line of stable equilibrium states that exist over a wide temperature range. The three colored curves indicate non-equilibrium thermodynamic paths that start from initial states I$_i$, II$_i$, and III$_i$, and end at stable equilibrium states I$_f$, II$_f$, and III$_f$. Since temperature is only defined along the black curve of stable equilibrium states in Figure~\ref{Cyclo_Paths}, it is not useful for describing non-equilibrium states.  Instead, we classify non-equilibrium paths in terms of how the energy of the brush subsystem changes. When brush energy increases, it is accomplished by the flow of energy in a heat interaction from the reservoir into the brush. Thus, a brush whose energy increases along a path is ``heated'' by the reservoir.  Conversely, if the brush energy decreases, energy in a heat interaction necessarily leaves the brush and flows into the reservoir. Paths with decreasing energy are thus ``cooling'' processes.

The equation of motion is deterministic and does not introduce statistical error into the calculation of the descriptor property values. Thus the error-bars in the succeeding figures arise from the maximum and minimum range of values expected from the uncertainty in the degeneracies of Figure~\ref{DOS}.

The red curve (Path I) represents heating from a low initial temperature of 6 degrees on an arbitrary temperature scale~\footnote{The energy eigenstructure of the brush system includes only bonding contributions. Therefore, there are no kinetic energy contributions from thermal vibrations or translational motions. Consequently, the temperatures noted refer to an arbitrary absolute temperature scale but not Kelvin degrees.} to a higher reservoir temperature at 300 degrees. The blue path (Path II) represents net heating (i.e., both heating and cooling occur) to the same reservoir temperature but from an initial near-equilibrium state very far below the final temperature. The orange curve (Path III) corresponds to cooling from an initial non-equilibrium (and high-energy) state to equilibrium at a reservoir temperature of 5 degrees. These three thermodynamic paths are selected to highlight the range of kinetics that can be produced by various choices of non-equilibrium initial states.

\begin{figure}
\centering
\includegraphics[width=0.45\textwidth]{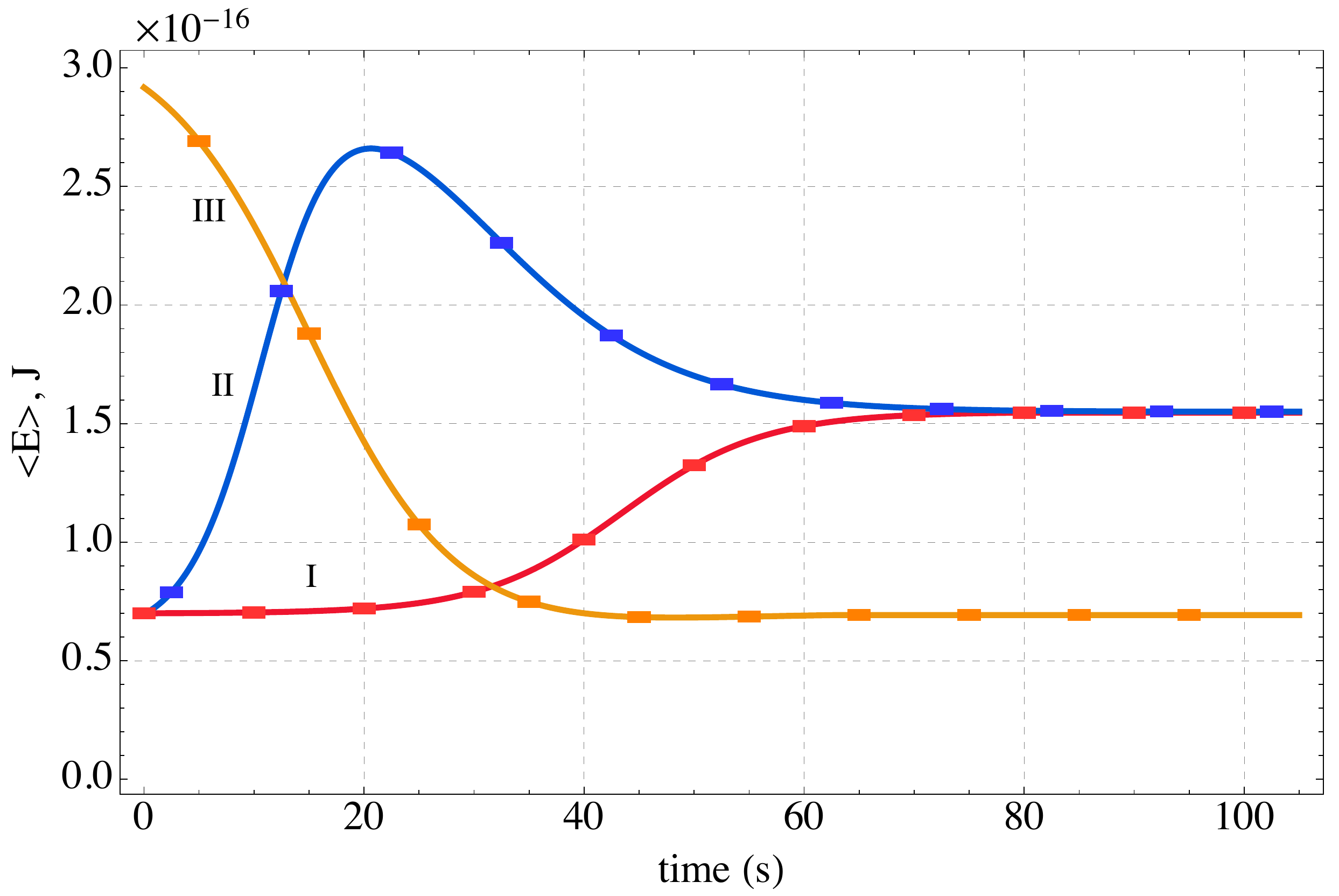}
\centering
\caption{Change in the energy of the system along the three non-equilibrium thermodynamic paths.}
\label{E_Change}
\end{figure}

\begin{figure}
\centering
\includegraphics[width=0.45\textwidth]{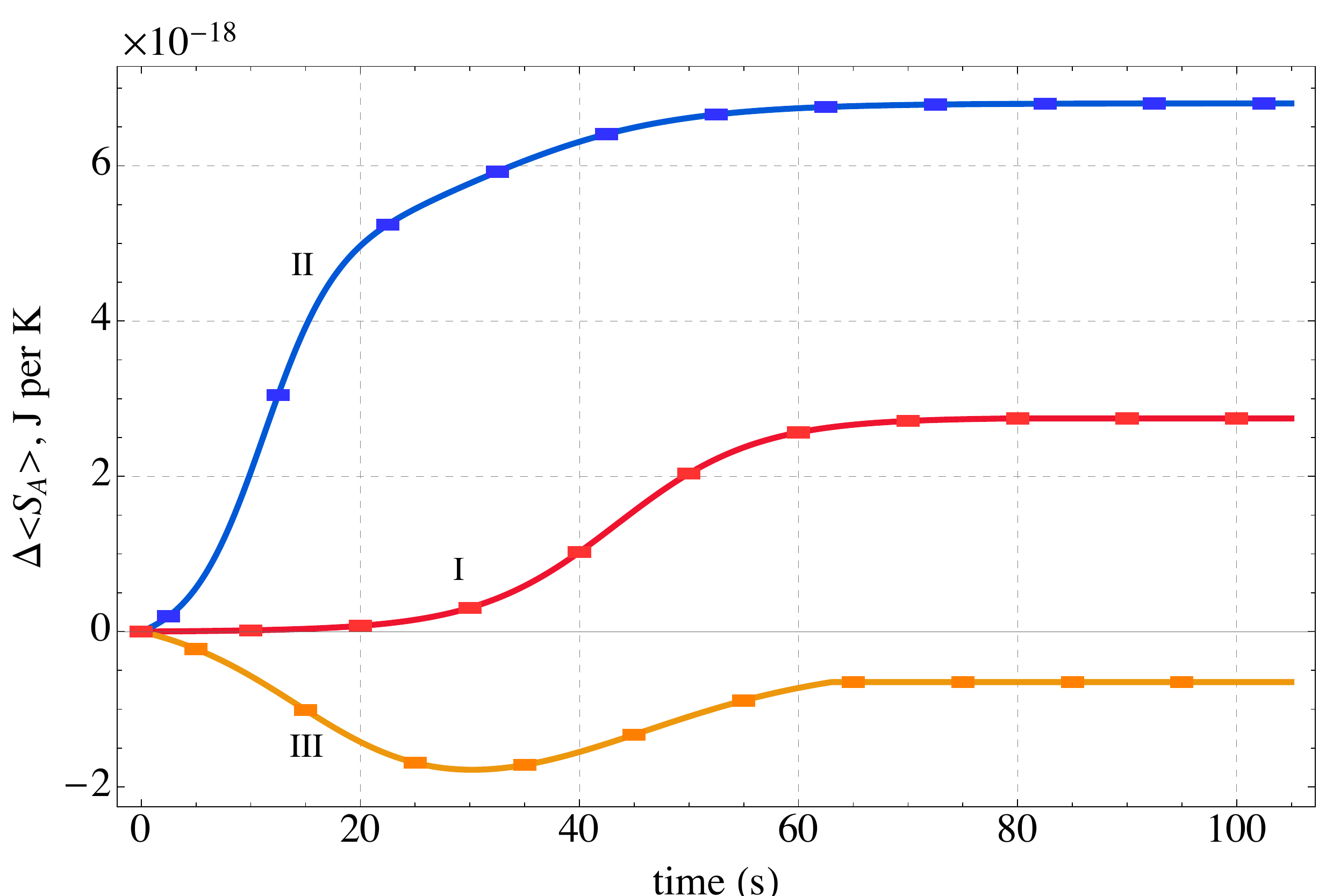}
\centering
\caption{Entropy change, $\Delta \langle S_A \, \rangle$, of the brush (subsystem $A$) for the three non-equilibrium paths.  The entropy change includes contributions from energy exchange with the reservoir as well as entropy generated within subsystem $A$.}
\label{Delta_SA}
\end{figure}

\begin{figure}
\centering
\includegraphics[width=0.45\textwidth]{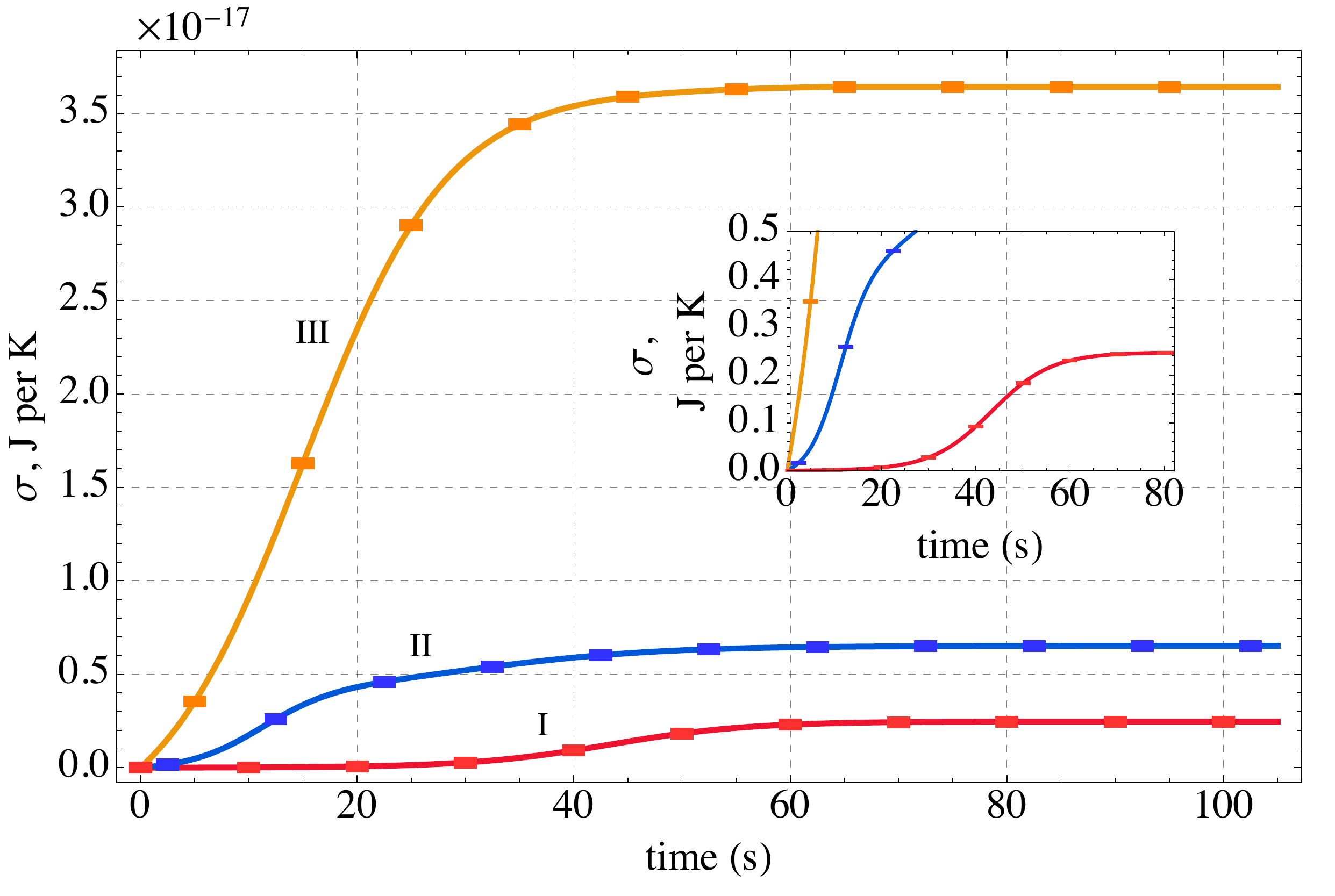}
\centering
\caption{Entropy generation for subsystem $A$ along the three non-equilibrium paths.}
\label{Sigma_Prod}
\end{figure}

\begin{figure}
\centering
\includegraphics[width=0.45\textwidth]{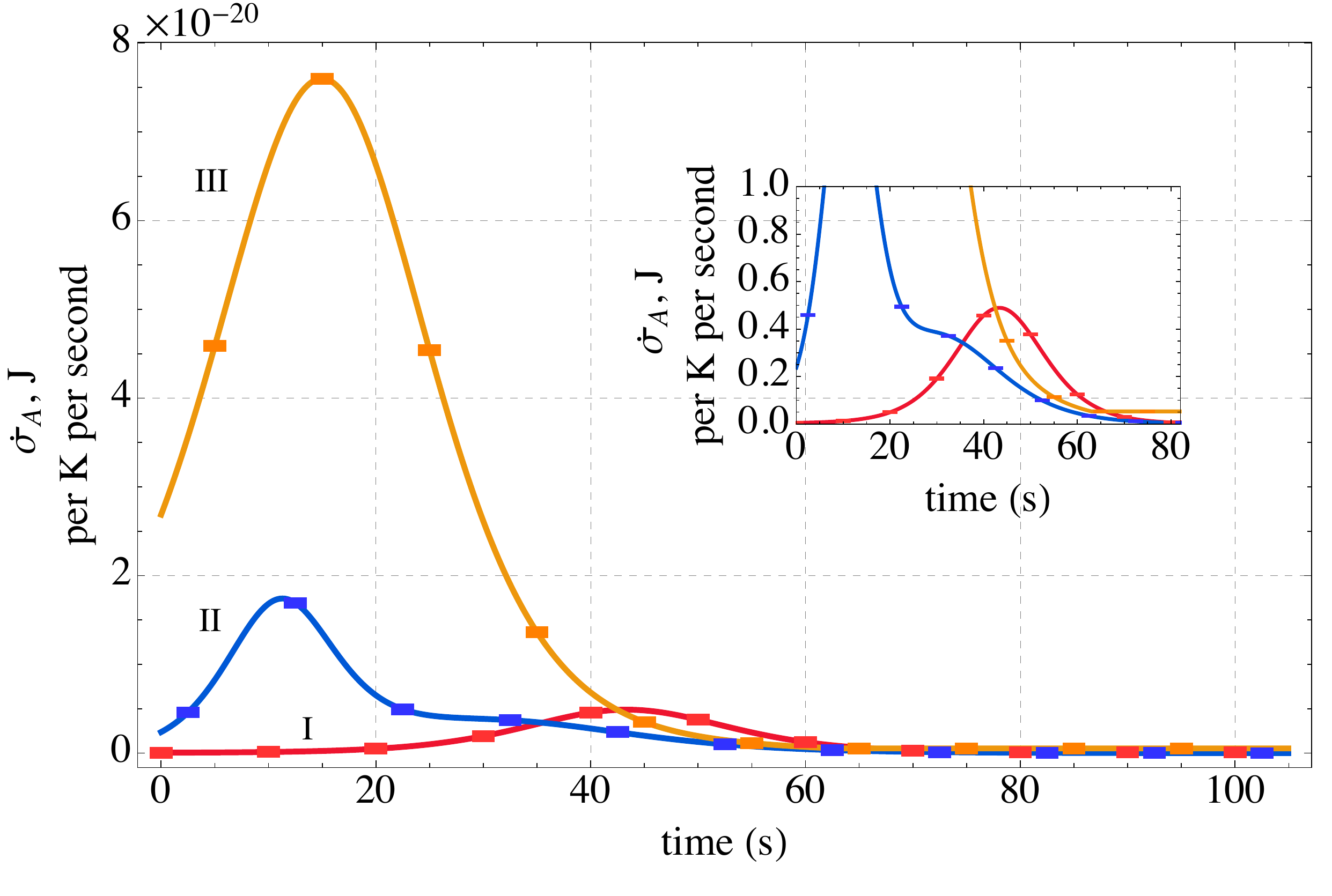}
\centering
\caption{Rate of entropy generation for subsystem $A$ along the three non-equilibrium paths.}
\label{Sigma_Dot}
\end{figure}

Consider first the nearly linear Path I of Figure~\ref{Cyclo_Paths}. As energy in a heat interaction flows from the high-temperature reservoir into subsystem $A$ (the brush), its energy increases monotonically. This increase is evident from the red curve in Figure~\ref{E_Change}, which shows the subsystem's energy as a function of time for each path. The rate of change in subsystem $A$'s energy (i.e., the energy flow into the brush) starts from zero, accelerates, and then slows down as the system approaches its final stable equilibrium state. 

In the case of Path II shown in Figure~\ref{Cyclo_Paths}, the energy of the brush rises to a maximum value near $1.6 \times 10^3$ J/mol (at a time of about $t=20$ seconds in Figure~\ref{E_Change}) and then decreases to its final value in mutual stable equilibrium with the high-temperature reservoir. As subsystem $A$'s energy increases and then decreases, the brush undergoes heating and then cooling along Path II. 

For Path III, subsystem $A$ starts in an initial state with a relatively high energy ($1.75 \times 10^3$ J/mol) that decreases smoothly as energy is transferred to the reservoir. Eventually, the system reaches equilibrium at $\langle E \rangle$ = $0.4 \times 10^3$ J/mol and $\langle S \rangle$ = 372 J/mol K. 

To gain insight into the behavior of the brush energy along the three paths, it is necessary to examine the evolution of the entropy. The SEAQT equation of motion determines each path by following the steepest ascent in entropy through state space. The stable equilibrium state is reached when the entropy of the closed composite system, consisting of subsystem $A$ (the brush) and the thermal reservoir, increases to the maximum value. However, as the brush is an open subsystem that exchanges energy with the reservoir, the entropy change of the brush does not necessarily simply increase. It is helpful to track the change in brush entropy, denoted by $\Delta \langle S_A \rangle$. There are three ways in which it changes: it decreases when energy flows out of subsystem $A$ to the reservoir, it increases when energy flows into subsystem $A$ from the reservoir, and it also increases when entropy is produced within subsystem $A$ as energy is irreversibly redistributed among the eigenlevels of the brush. We refer to this last contribution as the entropy generation, $\sigma_A$, where $\sigma_A \ge 0$. The equality in this expression holds for the special case of reversible or quasi-equilibrium paths. The time rate of change of energy generation is the energy generation rate, denoted by $\dot{\sigma}_A$. The quantities $\Delta \langle S_A \rangle$ and $\dot{\sigma}_A$ can be determined from the expected entropy, $\langle S_A \rangle$, the energy flow into and out of the reservoir calculated from the energy $\langle E_A \rangle$, and the reservoir temperature. Figure~\ref{Delta_SA} shows the entropy change of the brush, $\Delta \langle S_A \rangle$, along the three kinetic paths, 
while the entropy generation of the brush, $\sigma_A$, is presented in Figure~\ref{Sigma_Prod}.  The entropy generation {\em rate} of the brush, $\dot{\sigma}_A$, is the time derivative of the entropy generation, and it is proportional to the steepest entropy ascent at each time; this quantity is presented in Figure~\ref{Sigma_Dot}. 

Along Path I (the red curves in Figures~\ref{Delta_SA}, \ref{Sigma_Prod}, and \ref{Sigma_Dot}), the entropy change of the brush, $\Delta \langle S_A \rangle$, increases smoothly because it has two positive contributions along its entire path: ({\em i\/}) energy flows into the brush during heating (the red curve in Figure~\ref{E_Change} shows the brush energy increasing) and ({\em ii\/}) entropy is generated (red curve in Figures~\ref{Sigma_Prod} and \ref{Sigma_Dot}) by the redistribution of energy among the eigenlevels of the brush. These two positive contributions cause $\Delta \langle S_A \rangle$ to increase smoothly along Path I.

The trends in the entropy change of subsystem $A$ along Paths II and III are more complex. While steepest entropy ascent ensures that the entropy of the closed system increases monotonically to a maximum (evident in Figure~\ref{Sigma_Prod}), the entropy change of the brush subsystem can either increase or decrease depending on the relative magnitudes of energy flow and entropy generation. For the blue Path II, energy flows out of the brush during the cooling portion of the path (Figure~\ref{E_Change}), but the net entropy change in Figure~\ref{Delta_SA} is always positive because the positive entropy generation (Figure~\ref{Sigma_Prod}) more than compensates for the entropy decrease from cooling. On the other hand, for the orange Path III, the gradual loss of energy (Figure~\ref{E_Change}) from subsystem $A$ during cooling reduces the brush entropy more than entropy generation increases it, resulting in a negative net entropy change, $\Delta \langle S_A \rangle$.

Comparing the entropy generation, $\sigma_A$, for the three paths in Figure~\ref{Sigma_Prod}, it is evident that the entropy generation for Path I, which is closest to equilibrium of the three paths, is the smallest and lags behind that of Paths II and III. In addition, both of the latter two paths have entropy generation rates (Figure~\ref{Sigma_Dot}) that peak much higher than that of Path I. In fact, the entropy generation rate for the cooling case (Path III) is much higher than that for either of the other two paths, suggesting that it is the furthest from equilibrium relative to the reservoir with which the system interacts.

\subsubsection{Evolution of the Structural Parameters and System Microstructures}

\begin{figure}
\centering
\includegraphics[width=0.45\textwidth]{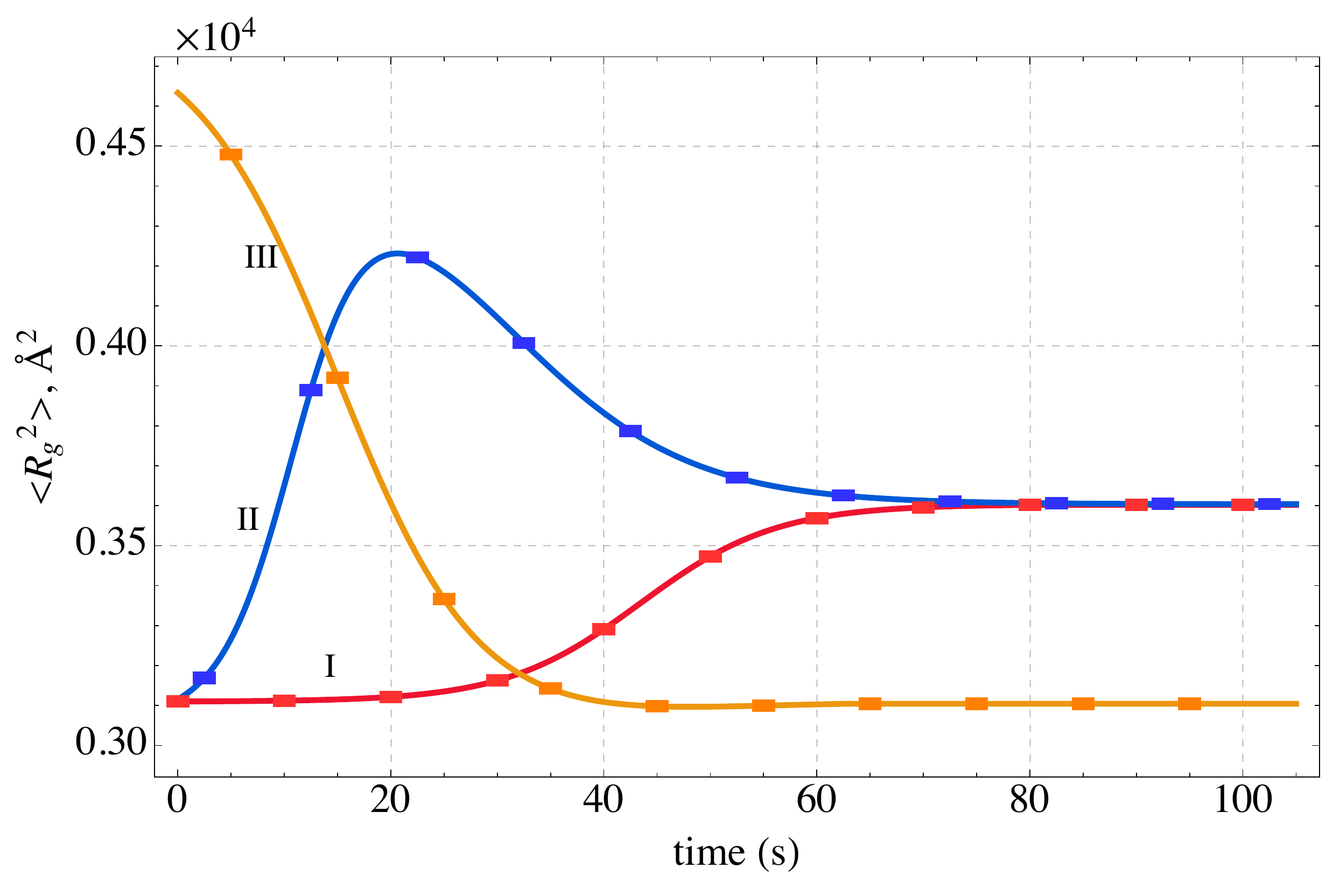}
\centering
\caption{Time evolution of the radius of gyration for the three non-equilibrium thermodynamic paths.}
\label{Cyclo_Rg}
\end{figure}

\begin{figure}
\centering
\includegraphics[width=0.45\textwidth]{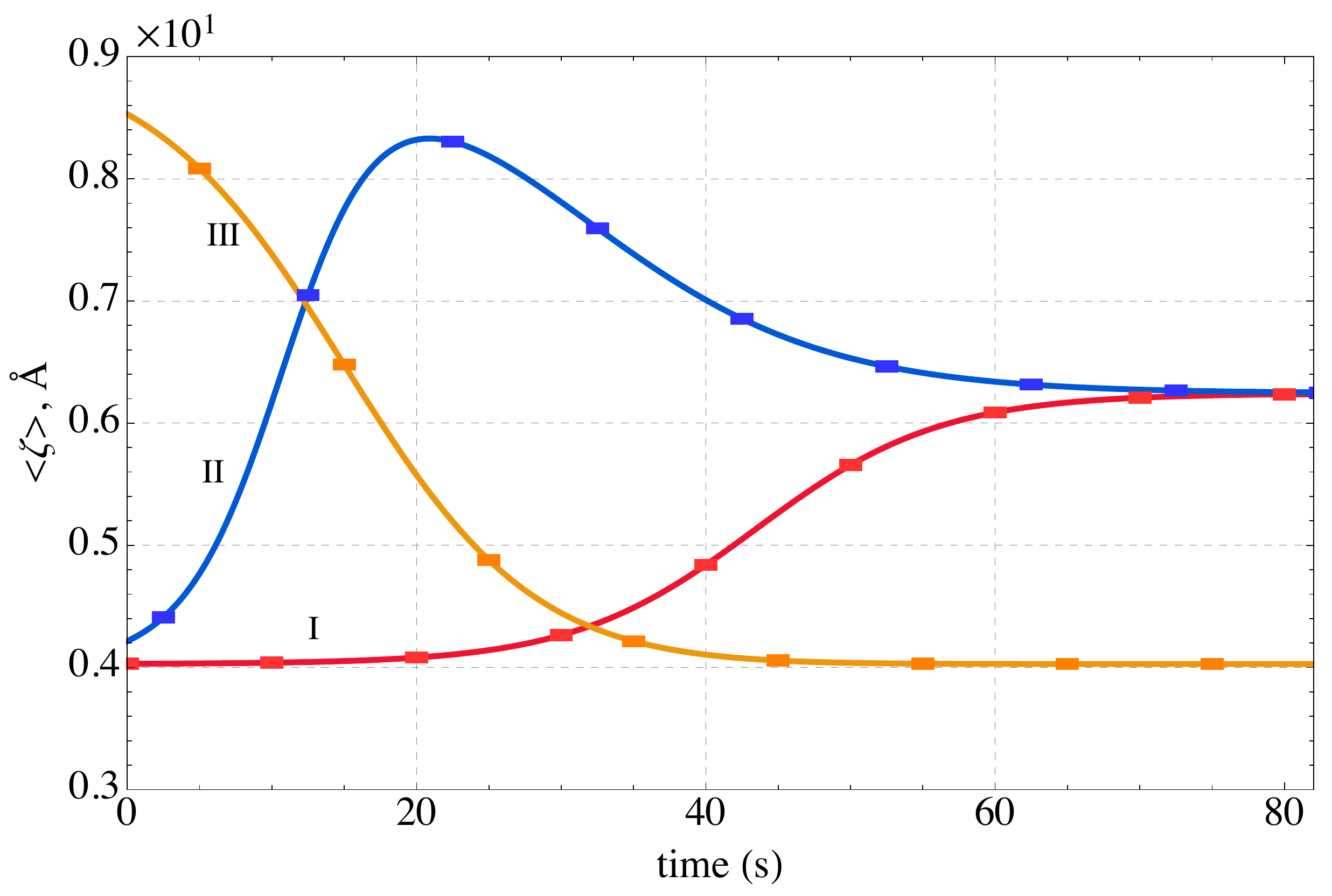}
\centering
\caption{Time evolution of the tortuosity for the three non-equilibrium thermodynamic paths.}
\label{Cyclo_t}
\end{figure}

In addition to determining changes in system energy and entropy, the occupation probabilities predicted by the SEAQT equation of motion also dictate the evolution of the structural properties and microstructure over time. Figures~\ref{Cyclo_Rg} and~\ref{Cyclo_t} show the time dependence of the radius of gyration and tortuosity, respectively, for each of the thermodynamic paths of Figure~\ref{Cyclo_Paths}. Both the radius of gyration and tortuosity follow the same trends as the expected energy in Figure~\ref{E_Change}. Increasing the system energy generally results in an increase in the radius of gyration and tortuosity, indicating that the chains in the brush extend and the number of bends along the chains increases. Conversely, decreasing the energy of the brush decreases the radius of gyration and tortuosity. Notably, the changes in these physical properties are most significant along paths that are farthest from equilibrium (Paths II and III), while Path I, being closest to equilibrium, exhibits the least change in the radius of gyration and tortuosity.  This is most evident when comparing Paths I and II. Despite the fact that these paths have very similar initial and final states, the variation in tortuosity along non-equilibrium Path II is about two times larger than along non-equilibrium Path I (Figure~\ref{Cyclo_t}).  A similar tendency can be seen for the radius of gyration and for the densities of polystyrene and solvent (discussed below), although these properties vary less than the tortuosity. Clearly, a physical property varies more along a path the farther the path is from equilibrium.

\begin{figure}

\begin{center}

\begin{subfigure}{0.45\textwidth}
    \includegraphics[width=\linewidth]{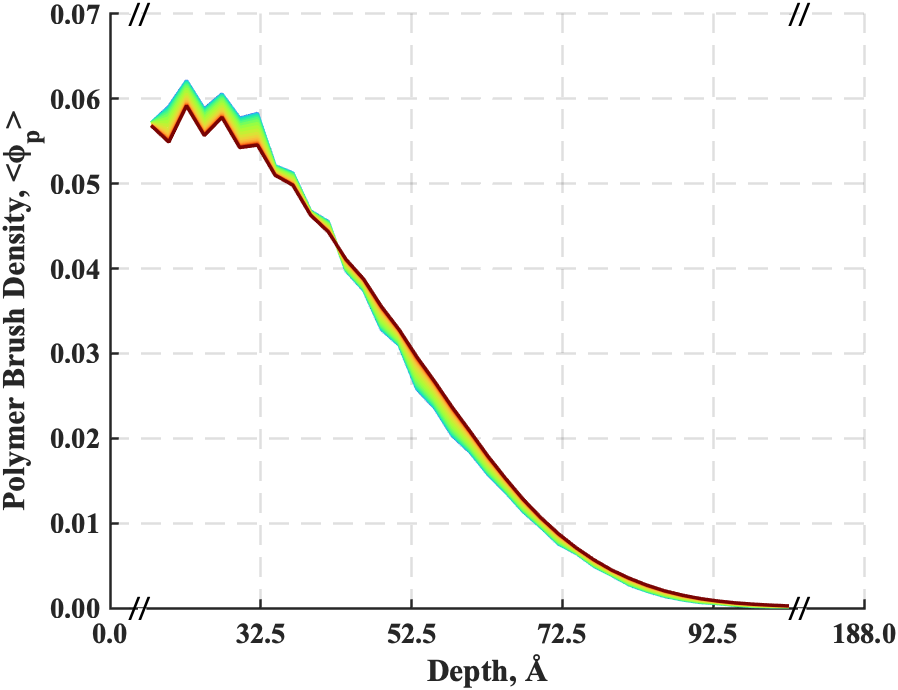}
    \caption{\vspace*{5mm}} 
    \label{fig:PolyStyreneDensityA}
  \hspace*{\fill}   
    \includegraphics[width=\linewidth]{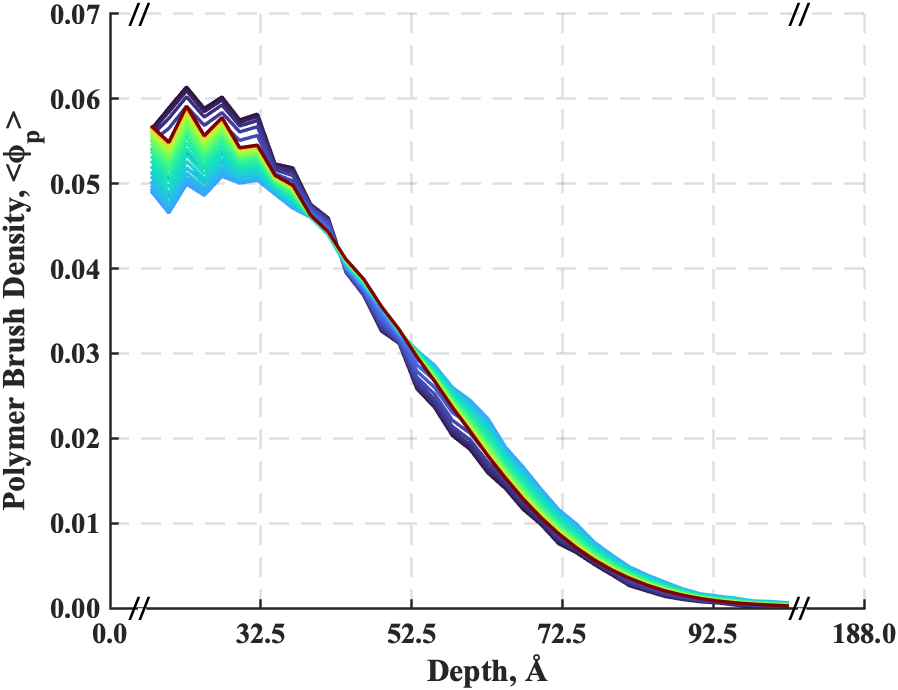}
    \caption{\vspace*{5mm}} 
    \label{fig:PolyStyreneDensityB}
  \hspace*{\fill}   
    \includegraphics[width=\linewidth]{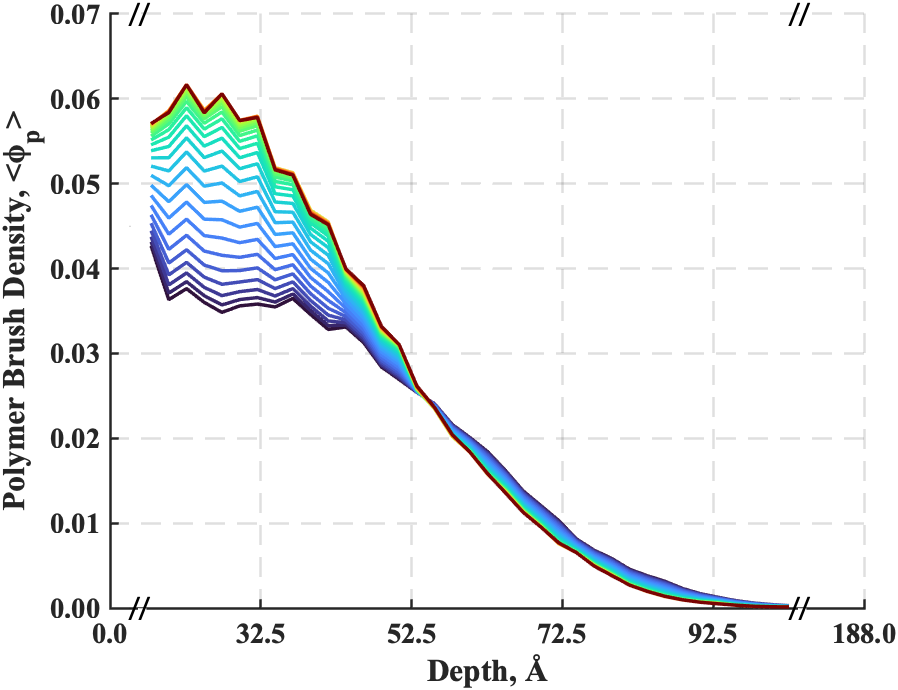}
    \caption{\vspace*{5mm}} 
    \label{fig:PolyStyreneDensityC}
  \end{subfigure}%
  \hspace*{\fill}

	\caption{Polystyrene density profiles in the polystyrene-cyclohexane system at various times for (a) Path I, (b) Path II, and (c) Path III.  The initial density profile is dark blue, the final density profile at stable equilibrium is dark red, and density profiles at intermediate times are represented by the hues between these extremes. The brush depth is plotted from $0$ (the grafting surface) to about 120 $\textrm{\AA}$ where the brush density declines to zero.}
		\label{Cyclo_DP}
	\end{center}
\end{figure}

\begin{figure}
\centering
\includegraphics[width=.45\textwidth]{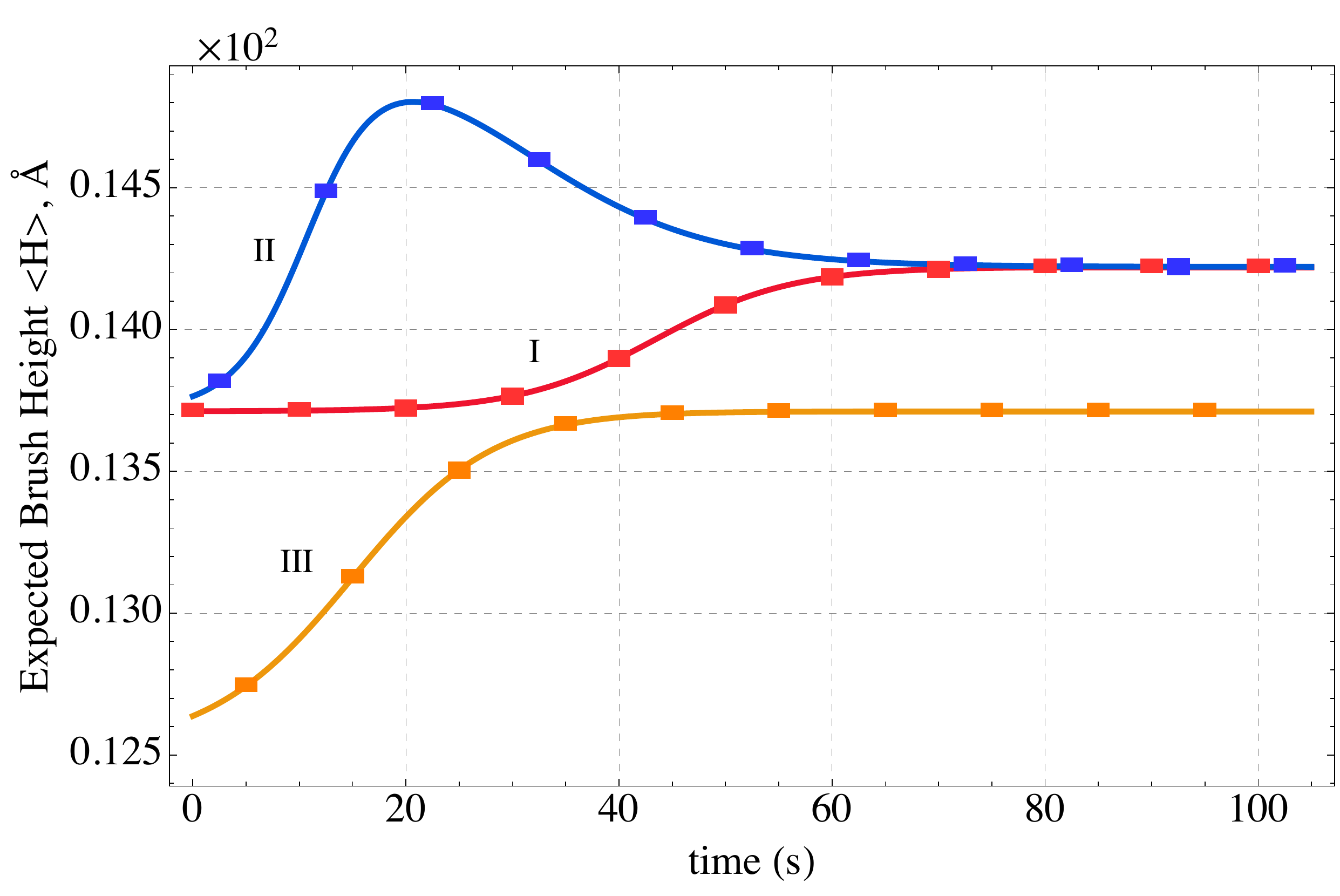}
\includegraphics[width=.45\textwidth]{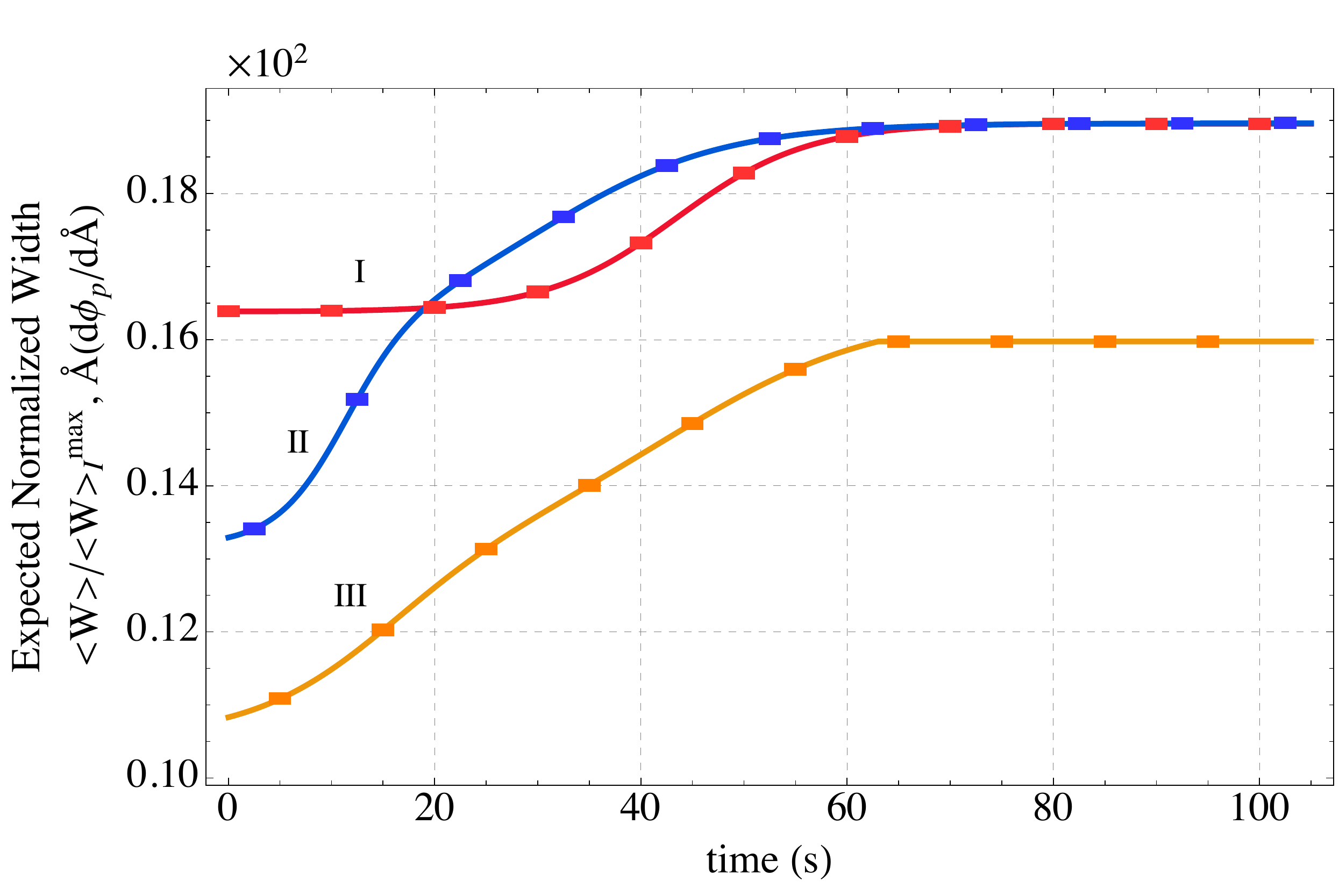}
\centering
\caption{Expected brush height and expected normalized width calculated form density profiles in Figure~\ref{Cyclo_DP} for the three non-equilibrium thermodynamic paths (indicated by color and the numbers I, II, and III); normalization of the expected brush width utilizes the maximum value of 44.5~\AA ~reached by Paths I and II.}
		\label{Cyclo_Brush_Param_Height}
\end{figure}

\begin{figure}
	\begin{center}
 \begin{subfigure}{0.45\textwidth}
    \includegraphics[width=\linewidth]{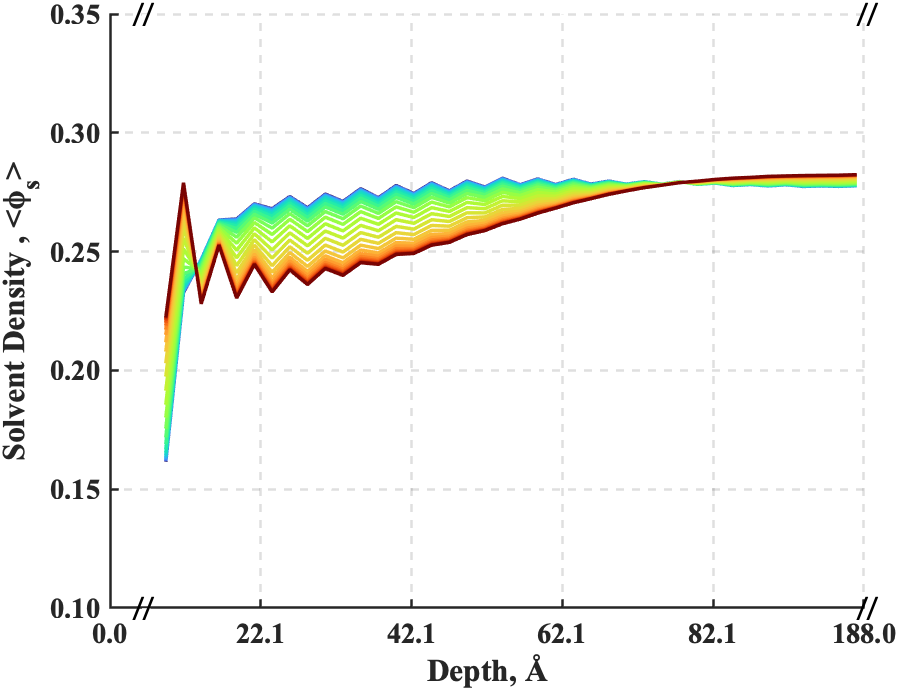}
    \caption{\vspace*{5mm}} 
    \label{fig:CycloDensityA}
  \hspace*{\fill}   
    \includegraphics[width=\linewidth]{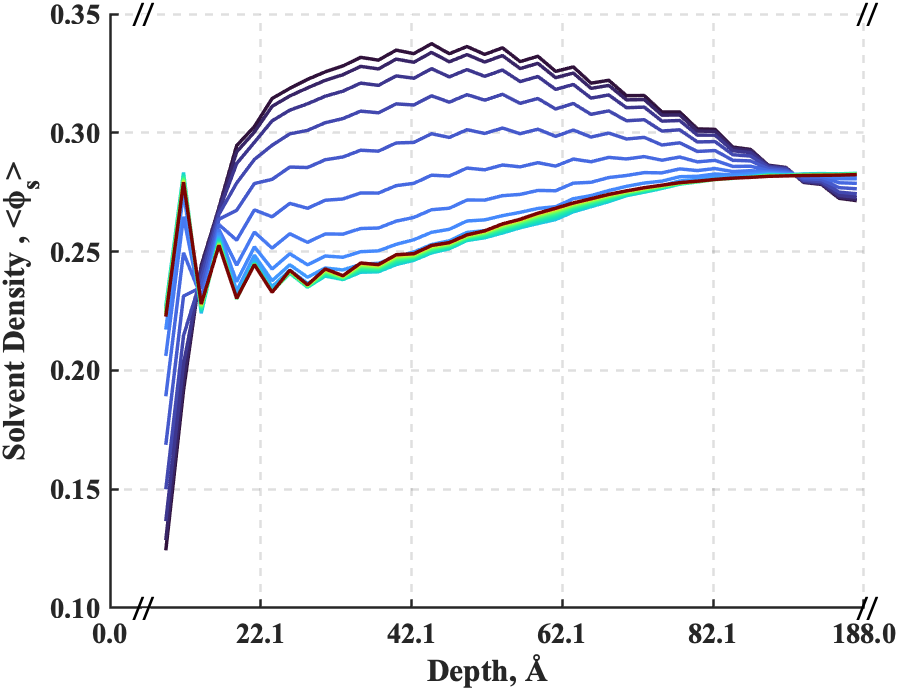}
    \caption{\vspace*{5mm}} 
    \label{fig:CycloDensityB}
  \hspace*{\fill}   
    \includegraphics[width=\linewidth]{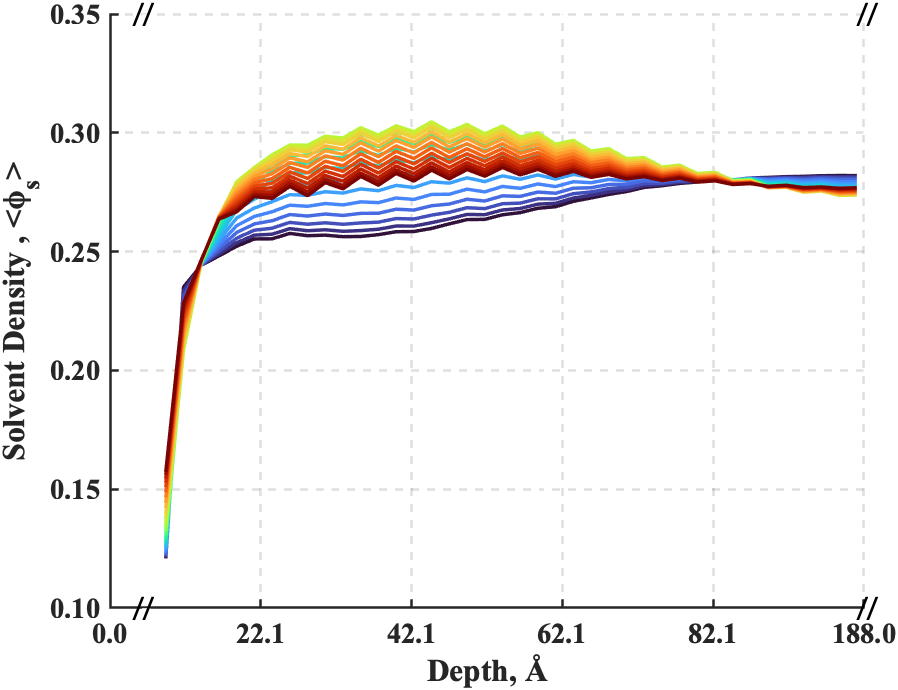}
    \caption{\vspace*{5mm}} 
    \label{fig:CycloDensityC}
  \end{subfigure}%
  \hspace*{\fill}
		\caption{Cyclohexane solvent density profiles in the polystyrene-cyclohexane system for (a) Path I, (b) Path II, and (c) Path III. The initial density profile is dark blue, the final density profile at stable equilibrium is dark red, and density profiles at intermediate times are represented by the hues between these extremes. The brush extends from a depth of 0 at the grafting plane to a depth of approximately 120 \AA ~beyond which there are no polystyrene chains and the solvent density is essentially constant.}
		\label{Cyclo_DPS}
	\end{center}
\end{figure}

The density profiles of polystyrene at various times for the three paths are presented in Figure~\ref{Cyclo_DP}. The density profile at each time is represented by a colored curve, where the initial density profile is dark blue, and the colors of the subsequent profiles shift towards dark red (the final stable equilibrium profile). The closely-spaced curves correspond to slowly changing density profiles, whereas the widely-spaced curves signify density profiles that evolve more rapidly. Missing profile colors are obscured by subsequent profiles that overlap them. 

Overall, the density profiles are consistent with the trends in brush length that are suggested by the predicted changes in the radius of gyration and tortuosity (Figures~\ref{Cyclo_Rg} and~\ref{Cyclo_t}).  Changes in brush density are most noticeable at small depths from the grafting surface at a depth of zero; beyond about $50 $\AA, the density does not change much with time. Also, the brush density near the grafting surface and the density closer to the free ends of the brush ($\gtrsim 50 $\AA) change with time in opposite directions. 

The profiles in Figure~\ref{Cyclo_DP} for Path I exhibit a slight but steady decline in the maximum density (at a depth $\sim 20 $\AA) over time and a slight reduction in the maximum density gradient as the system approaches stable equilibrium. These changes in the density profiles along Path I correspond to an increase in the brush height and width, two parameters that are plotted in Figure~\ref{Cyclo_Brush_Param_Height} \footnote{The width is normalized using the maximum value of 44.5 \AA ~from Paths I and II; this is intended as an indirect analog to the $\theta$-solvent condition used for normalizing experimental results found in the literature. This condition for polymers occurs at about 300 K for which there is no direct analog in this paper since the 300 degree units of the absolute temperature of the high-temperature reservoir used here lacks, as discussed earlier, contributions from the vibrational and translational modes of the polymers. Nonetheless, it is assumed that using the maximum width of the first two paths, which occurs at 300 degrees, permits a qualitative comparison with the experimental data. Indeed this appears to be the case, showing good qualitative agreement as seen in Figure~\ref{Cyclo_Brush_Param_Height} where the expected normalized width ranges from about 0.57 to 1.0. while the experimental results in Karim $et. \; al$.~\cite{Karim1994} range from about 0.57 to 1.1.}. Returning to the brush density profiles of Figure~\ref{Cyclo_DP}, the change in time along Path II is more complex than along Path I because the brush chains move differently in the length and width directions.  The brush height initially increases, reaches a maximum, then decreases with time (as seen in the blue Path II in Figure~\ref{Cyclo_Brush_Param_Height}), whereas the brush width steadily increases. Thus, the density near the grafting plane initially decreases quickly, then turns around and gradually increases. The density profile near the grafting plane along Path III steadily increases (the density profile near the free end decreases). These profile changes follow the monotonic increases in brush height and width along Path III of Figure~\ref{Cyclo_Brush_Param_Height}. The polymer density profiles change monotonically for Paths I and III, but not for Path II. As was the case for the tortuosity and radius of gyration along Paths II and III, the polymer density profiles, brush height, and brush width also change most noticeably along these paths.

The density profile of the {\em solvent} also can be tracked with time. Figure~\ref{Cyclo_DPS} presents the density profiles for cyclohexane at various times along the three thermodynamic paths. Just as for the corresponding polymer density profiles in Figure~\ref{Cyclo_DP}, the initial solvent density profile is dark blue and the colors of the subsequent profiles shift towards dark red (the final stable equilibrium profile). The most significant changes in solvent density take place near the middle of the brush.  The solvent density profile along Path I decreases smoothly with time as the brush is heated: departing from the initial blue profile slowly, accelerating, and finally approaching the equilibrium red profile slowly.  The solvent profile also decreases smoothly along Path II (also heating), but the decrease from the initial blue profile is quite rapid (widely separated blue profiles) and there is very little change during the final approach to the final red profile. The solvent density profiles along cooling Path III first increase then decrease.  The solvent density profiles change monotonically for Paths I and II, but not for Path III.   

\begin{figure}
	\begin{center}
 \begin{subfigure}{0.45\textwidth}
    \includegraphics[width=\linewidth]{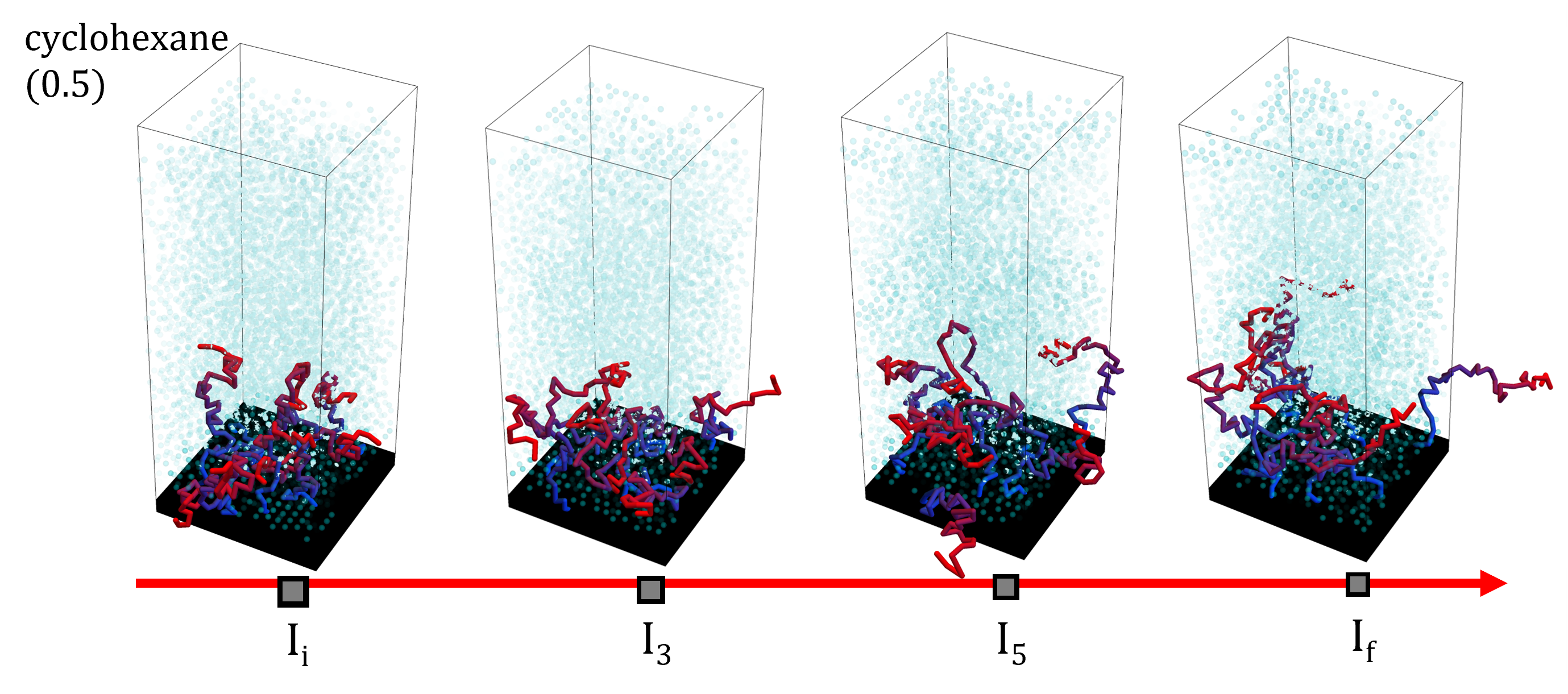}
    \caption{\vspace*{5mm}} 
    \label{fig:CyclohexaneMicroA}
  \hspace*{\fill}   
    \includegraphics[width=\linewidth]{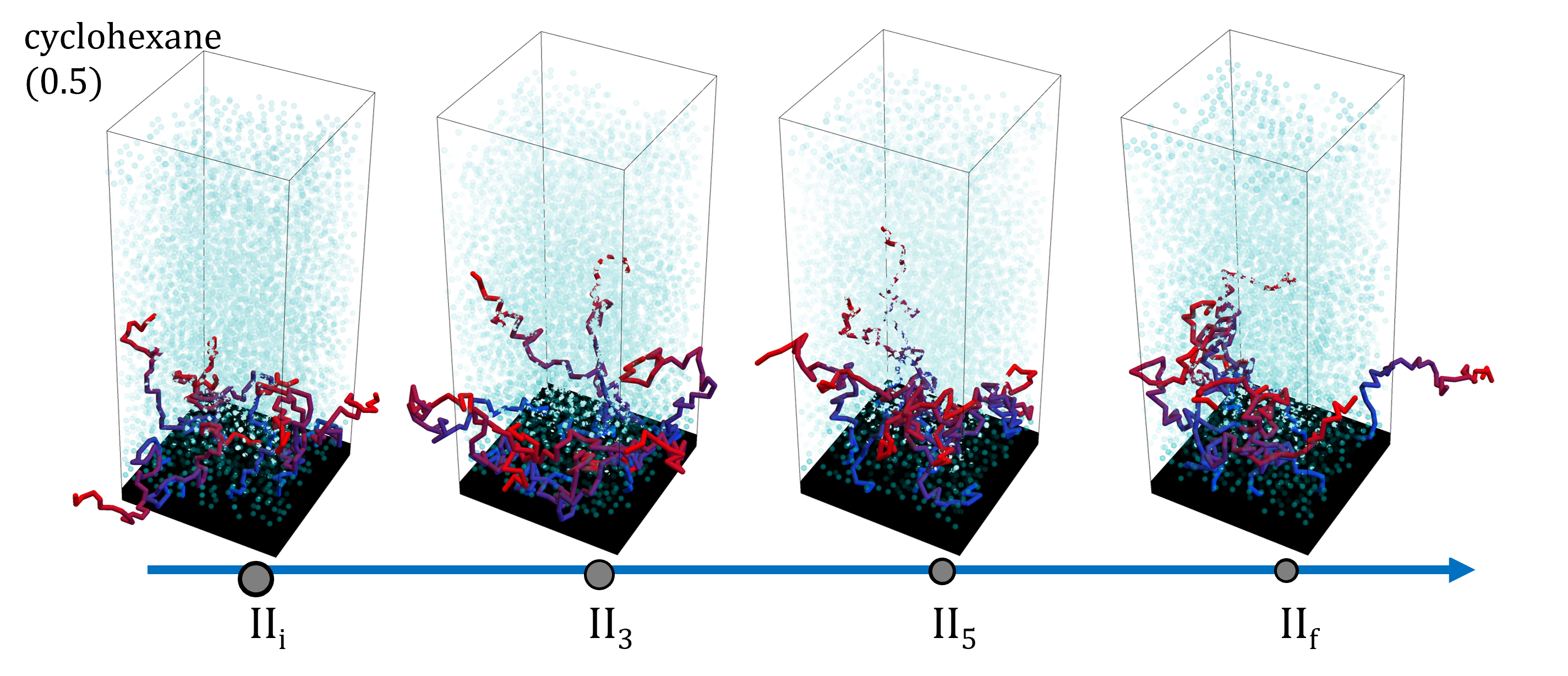}
    \caption{\vspace*{5mm}} 
    \label{fig:CyclohexaneMicroB}
  \hspace*{\fill}   
    \includegraphics[width=\linewidth]{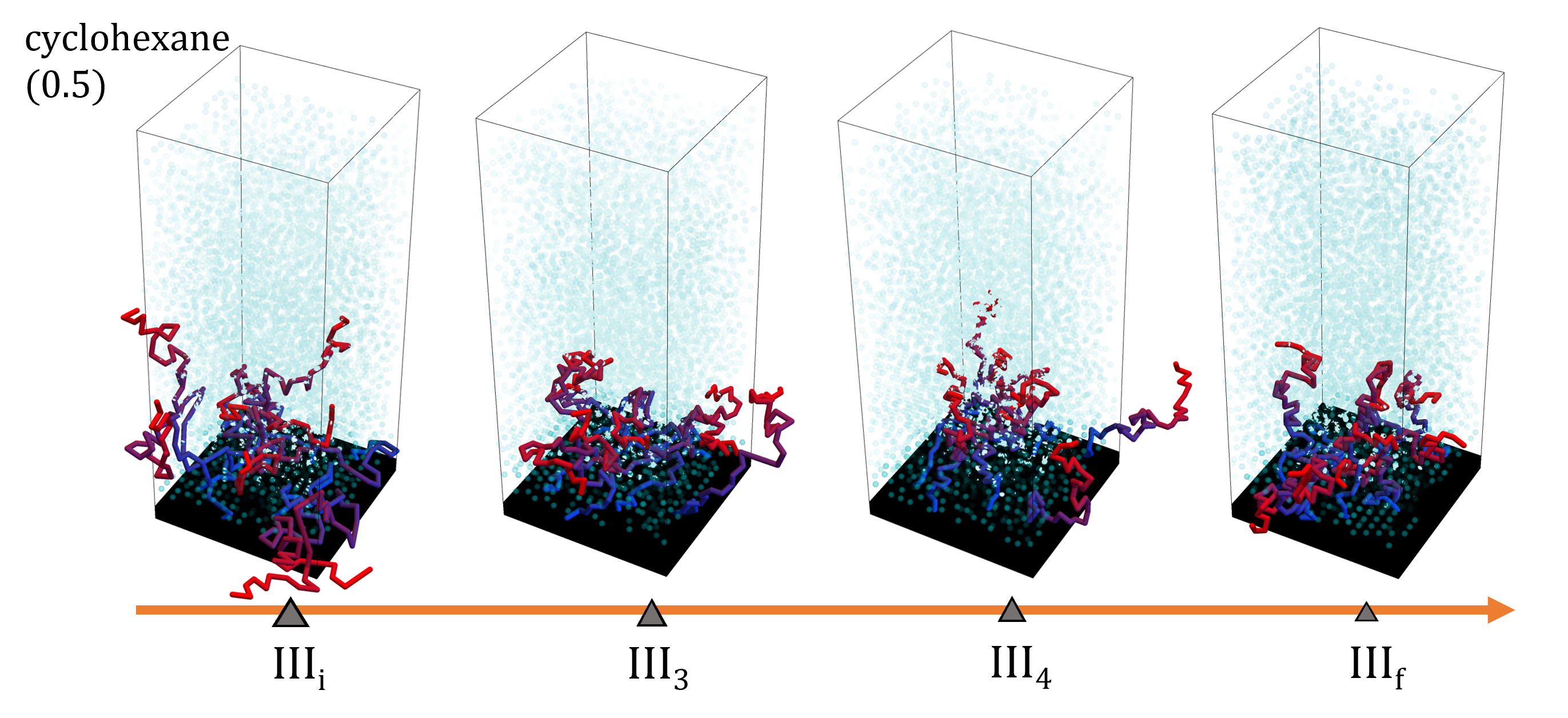}
    \caption{\vspace*{5mm}} 
    \label{fig:CyclohexaneMicroC}
  \end{subfigure}%
  \hspace*{\fill}
		\caption{Representative expectation microstructures for the brush system in cyclohexane along (a) Path I, (b) Path II, and (c) Path III. Polymer coloring is consistent with that in Figure~\ref{ExampleBrush} with the color for the polymer changing from blue at the grafting plane of the brush to red at the free end. The solvent is colored light blue with a slight change in opacity dependent on local solvent concentration.}
		\label{Cyclo_Rep_Micro}
	\end{center}
\end{figure}

Figure~\ref{Cyclo_Rep_Micro} displays representative expectation microstructures for Paths I, II, and III. Four microstructures are shown for each path: the initial conformation (i), two non-equilibrium conformations at intermediate times, and the final stable equilibrium conformation (f). Figure~\ref{Cyclo_Paths} shows the locations of these states along the thermodynamic paths. These microstructures were chosen from a set of microstructures identified during the process of determining the energy eigenstructure of the system using the Replica-Exchange Wang-Landau algorithm. The choice from this set was based on the expected structural parameter values found for a given state using the equation of motion.

One of the significant features of brush conformations is the difference in chain extension among the three paths. For Path I, starting from the lowest energy state of the chains's initial conformation, the polymer transitions from a clustered state to an increasingly extended dispersion as energy flows from the reservoir to the brush during heating. Path II starts from a less clustered state than Path I but quickly progresses to extended chains before settling into the same final equilibrium state as Path I. On the other hand, Path III starts with a high initial energy and extended chains, but as the brush loses energy during cooling, the conformations become progressively more clustered.

\subsection{Toluene solvent}
\subsubsection{Evolution of the Structural Parameters and System Microstructures}

To investigate how sensitive the approach is to the type of solvent used, we recalculated the energy eigenstructure and degeneracies for a polystyrene brush in toluene, as opposed to cyclohexane. We also tested the brush's responsiveness to different solvent characteristics by using both the negative and positive values for the Flory-Huggins parameter for polystyrene in toluene listed in Table~\ref{PairPotentials}. The purpose of this was to determine whether the density profiles we calculated, as well as those reported in the literature \cite{Karim1994}, are significantly affected by the energy interactions between the polymer and solvent under different conditions.

Now, since only equilibrium results are available in the literature, a relatively near-equilibrium path is generated here for both polymer-toluene systems (one with a positive and the other with a negative Flory-Huggins parameter) that mimic Path I from the polystyrene-cyclohexane system in Section~\nameref{subsec:CyclohexaneSolvent}. This is done by using the initial state and thermal reservoir for the toluene systems as was used for Path I in the cyclohexane solvent. This yielded the same type of linear paths as Path I in Figure~\ref{Cyclo_Paths}, but the energy and entropy values differ because the energy eigenstructures differ. The behaviors of the expected structural parameters (i.e., radius of gyration and tortuosity) for both toluene systems are very similar to those already shown for the cyclohexane system and, are thus, not shown. The behaviors of the polymer solvent density profiles for both toluene systems, on the other hand, are presented since these results can be compared with results found in the literature.

\begin{figure}
	\begin{center}

   \begin{subfigure}{0.45\textwidth}
    \includegraphics[width=\linewidth]{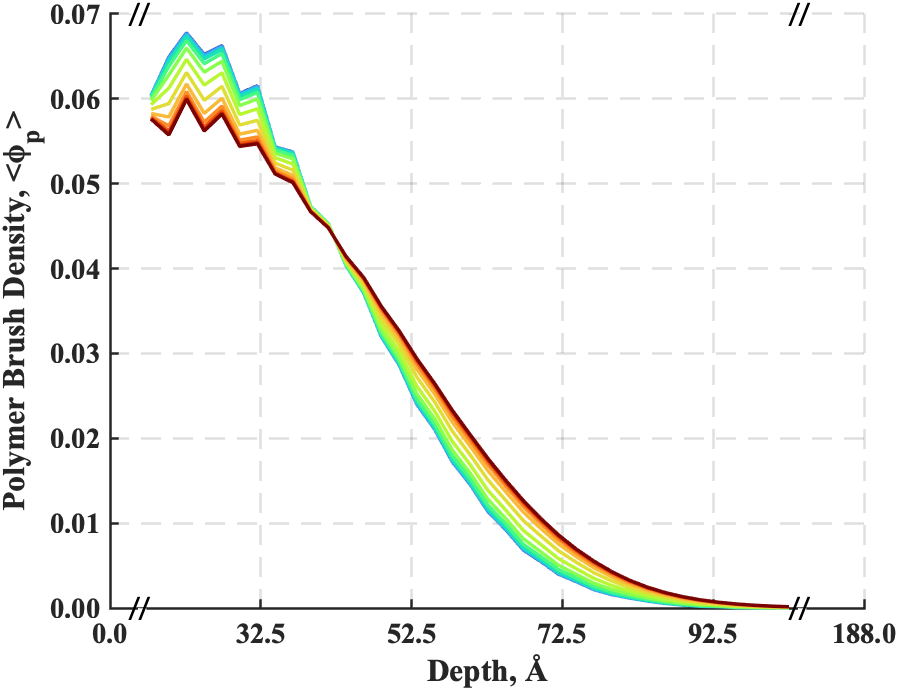}
    \caption{\vspace*{5mm}} 
    \label{fig:PSdensityA}
  \hspace*{\fill}   
    \includegraphics[width=\linewidth]{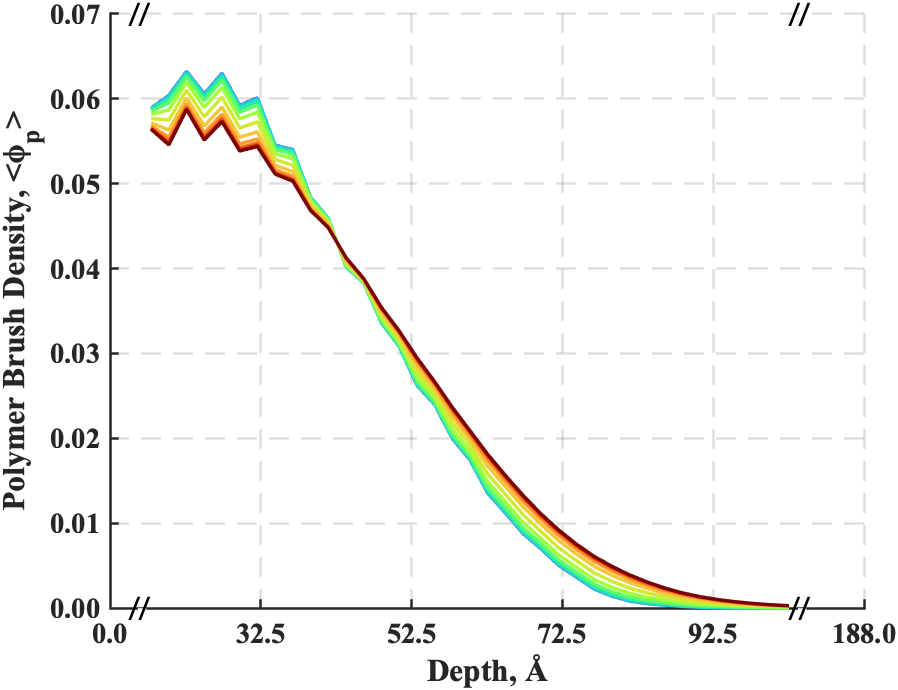}
    \caption{\vspace*{5mm}} 
    \label{fig:PSdensityB}
  \end{subfigure}%
  \hspace*{\fill}
		\caption{Polystyrene density profiles in the polystyrene-toluene system at various times along a path from an initial state at 5 degrees to a final stable equilibrium states at 300 degrees (similar to Path I of Figure~\ref{Cyclo_Paths}). Part (a) was calculated from an energy eigenstructure with a negative Flory-Huggins parameter and (b) from a landscape with a positive Flory-Huggins parameter. In both cases, the initial density profile is dark blue, the final density profile at stable equilibrium profile is dark red, and density profiles at intermediate times are represented by the hues between these extremes.}
		\label{Model_DP}
	\end{center}
\end{figure}
\begin{figure}
	\begin{center}
    \begin{subfigure}{0.45\textwidth}
    \includegraphics[width=\linewidth]{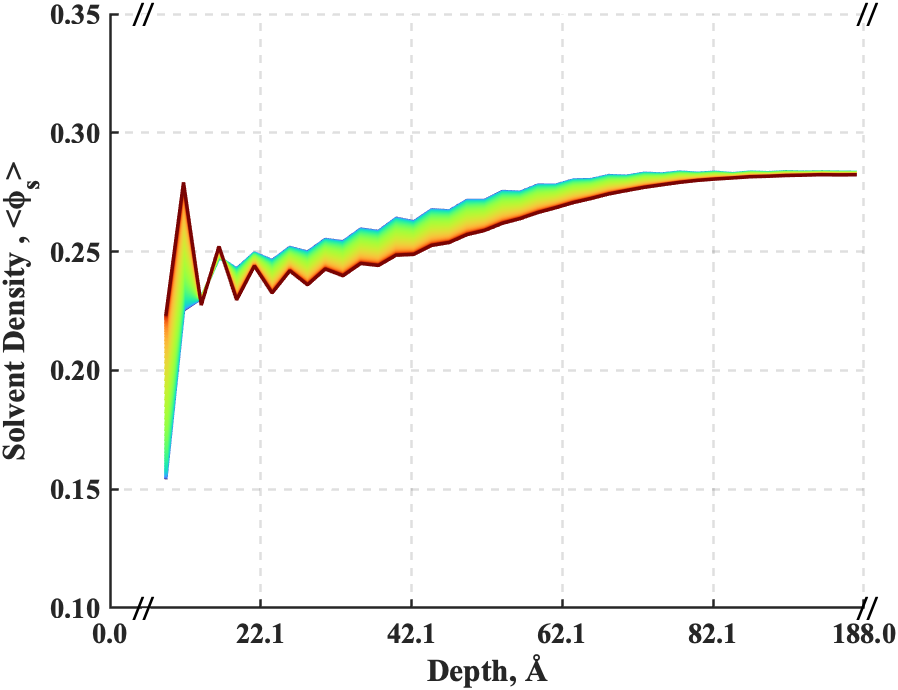}
    \caption{\vspace*{5mm}} 
    \label{fig:TolueneDensityA}
  \hspace*{\fill}   
    \includegraphics[width=\linewidth]{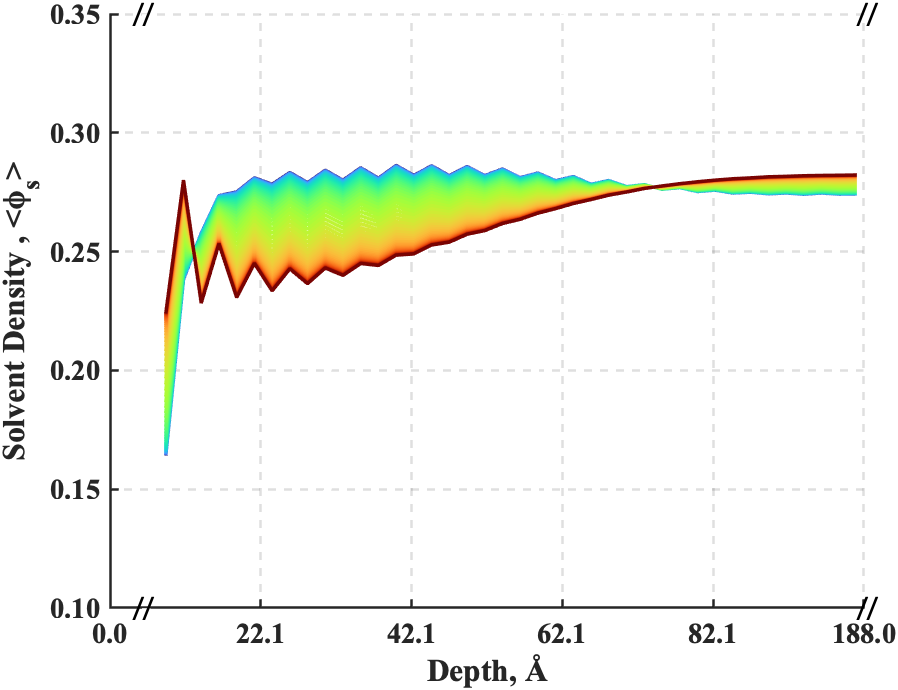}
    \caption{\vspace*{5mm}} 
    \label{fig:TolueneDensityB}
  \end{subfigure}%
  \hspace*{\fill}
		\caption{Toluene solvent density profiles in the polystyrene-toluene system at various times along a path from an initial state at 5 degrees to a final stable equilibrium states at 300 degrees (similar to Path I of Figure~\ref{Cyclo_Paths}). Part (a) was calculated from an energy eigenstructure with a negative Flory-Huggins parameter and (b) from a landscape with a positive Flory-Huggins parameter. In both cases, the initial density profile is dark blue, the final density profile at stable equilibrium profile is dark red, and density profiles at intermediate times are represented by the hues between these extremes.}
		\label{Model_DPS}
	\end{center}
\end{figure}

Figure~\ref{Model_DP} shows the polymer brush density profiles for polystyrene in both polymer-toluene systems. The final equilibrium polymer density profile is nearly identical when calculated with a negative or a positive Flory-Huggins parameter. Compared to the profiles for the polymer-cyclohexane system along Path I in Figure~\ref{Cyclo_DP}, the toluene systems contract slightly more from the initial state. However, despite the substantially different parameter values used to generate the energy eigenstructures (as shown in Table~\ref{PairPotentials}), the profiles for polystyrene in toluene with the positive Flory-Huggins parameter (bottom part of Figure~\ref{Model_DP}) are quite similar to the Path I profiles for the polymer-cyclohexane system (Figure~\ref{Cyclo_DP}).

In contrast to the polymer density profiles, the solvent density profiles for toluene were found to be sensitive to the sign of the Flory-Huggins parameter, as depicted in Figure~\ref{Model_DPS}. Once again, the positive Flory-Huggins polystyrene-toluene system (displayed in the bottom figure) was very similar to the solvent density profiles of Path I in the polymer-cyclohexane system (Figure~\ref{Cyclo_DPS}), except for a minor decrease in the depth of the initial reduction in density for the initial states. Changes in the solvent density profiles appear to be primarily influenced by the relative values of the component interactions of the Flory-Huggins parameter, i.e., when the solvent self-interaction is greater than the mixed interaction, the system displays little qualitative difference between the curves over a broad range of energies and entropies. Conversely, when the polymer-solvent interaction dominates, the solvent profile shows a significant inclination to infiltrate further into the brush at low energies.

\begin{figure}
	\begin{center}

      \begin{subfigure}{0.45\textwidth}
    \includegraphics[width=\linewidth]{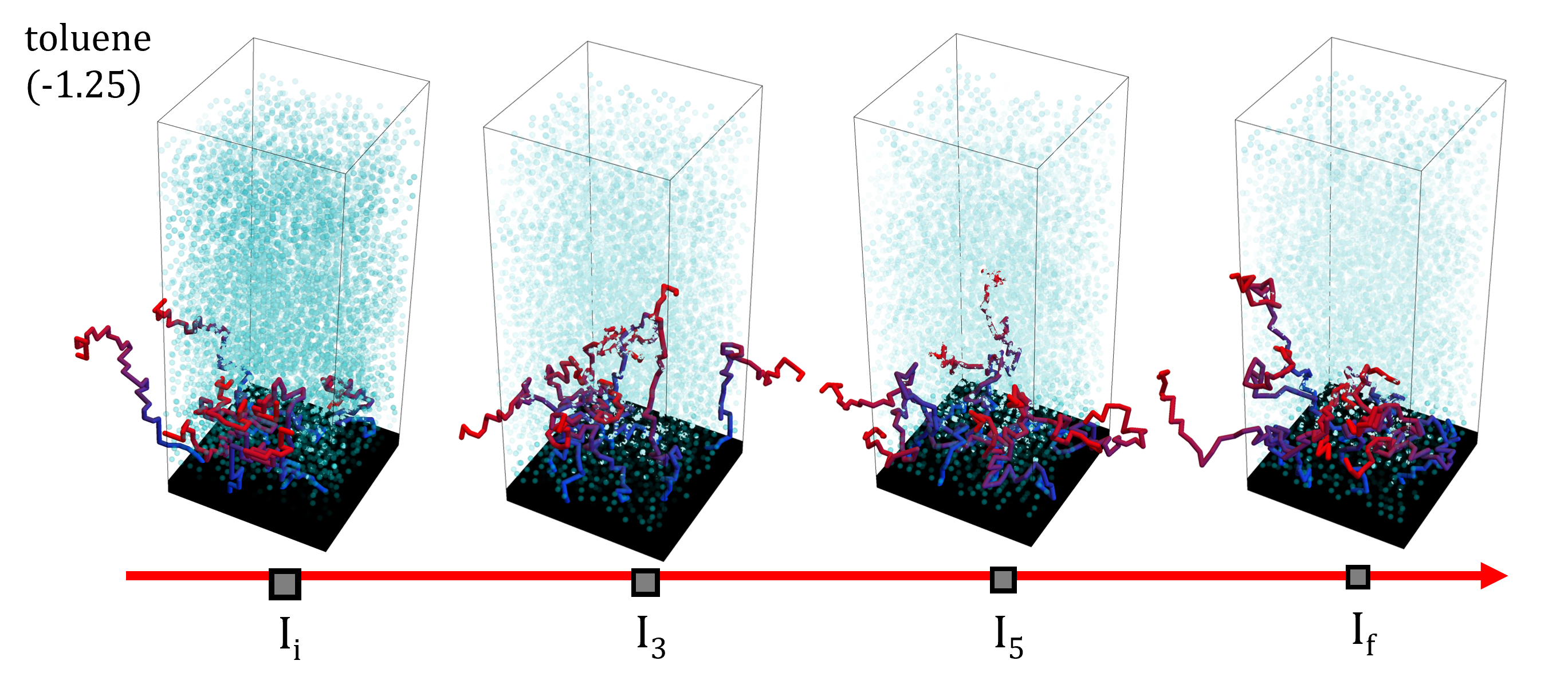}
    \caption{\vspace*{5mm}} 
    \label{fig:FLmicroA}
  \hspace*{\fill}   
    \includegraphics[width=\linewidth]{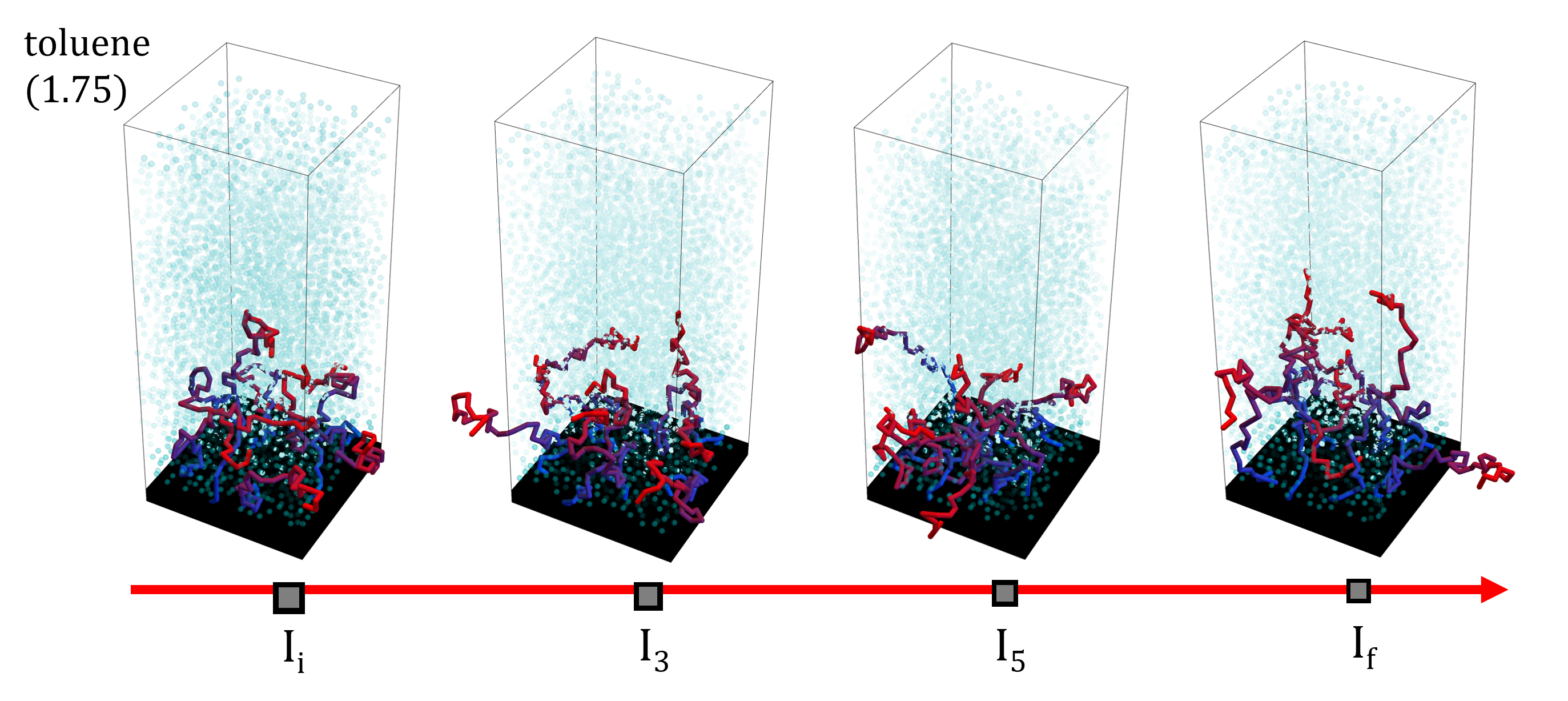}
    \caption{\vspace*{5mm}} 
    \label{fig:FLmicroB}
  \end{subfigure}%
  \hspace*{\fill}
        \caption{Representative expectation microstructures for the brush system in toluene at various times along a path from an initial state at 5 degrees to a final stable equilibrium state at 300 degrees (similar to Path I of Figure~\ref{Cyclo_Paths}). Part (a) was calculated from an energy eigenstructure with a negative Flory-Huggins parameter (-1.25) and (b) from a landscape with a positive Flory-Huggins parameter (1.75).}
		\label{Model_Rep_Micro}
	\end{center}
\end{figure}

Representative expectation microstructures for the two polymer-toluene systems are shown in Figure~\ref{Model_Rep_Micro}.  The top row of microstructures represents evolution for a negative Flory-Huggins parameter, and the bottom row corresponds to evolution for a positive Flory-Huggins parameter. The chain conformations at equivalent times along the paths of the two systems are essentially the same. The bottom row (positive Flory-Huggins parameter) in Figure~\ref{Model_Rep_Micro} is quite similar to the microstructures along Path I of the polymer-cyclohexane system in Figure~\ref{Cyclo_Rep_Micro}. In both systems, the microstructure begins in a state with a compact conformation and then proceeds through similar stages of chain extension.

\section{Discussion}
\label{section:Discussion}

The SEAQT equation of motion yields kinetic results that are expressed in non-dimensional times. To convert these times to the real times presented in Section~\nameref{sec:results}, Rouse dynamics were used (Equation~(\ref{tau})). For example, along Path I for the polymer-cyclohexane system, the $R_{g_\perp}^2$ changed by 150 \AA$^2$. The experimental value of the 3-dimensional diffusion coefficient for polystyrene is $2\times10^{-6} \;\text{cm}^2/\text{s}$, but since the brush chains are fixed to the grafting plane and lateral motion is restricted, this diffusion constant is reduced by a factor of $10^3$ to $2\times10^{-9}\;\text{cm}^2/\text{s}$. It is important to note that the change in the radius of gyration may not be strictly linear because the specific conformational relaxation involved in coiling and the potential overlap of chains impede motion. Therefore, an appropriate proportionality constant is needed to scale the results. Previous work~\cite{McDonald2023polymer} analyzed polymer dynamics using Rouse relations to translate SEAQT-derived dynamics of a simple polymer chain to real time. They used a proportionality constant of 2.35 $\times 10^7$, which scales the relaxation time to approximately 13.3 minutes. This scaled time aligns well with the temperature quenching conditions of experimental studies.

The three kinetic paths considered provide insight into the relative roles of entropy and energy on brush properties. Paths I, II, and III were determined by the steepest-entropy-ascent postulate underlying the equation of motion. The equation of motion effectively selected these paths to maximize the entropy generation rate curves of Figure~\ref{Sigma_Dot}, given the initial conditions and reservoir temperatures. A path through state space determines how physical properties change, raising the question of which state property (energy or entropy) serves as the most accurate proxy for physical properties.

Figures~\ref{Cyclo_Rg}, \ref{Cyclo_t}, \ref{Cyclo_DP}, and \ref{Cyclo_Brush_Param_Height} closely follow the time evolution of the brush's {\em energy}, Figure~\ref{E_Change}, indicating that the brush's energy is a more accurate proxy for most physical properties than its entropy. The one exception is the solvent density profile, Figure~\ref{Cyclo_DPS}, which tracks the time evolution of the brush's entropy, Figure~\ref{Delta_SA}. Evidently, the distribution of the solvent is more sensitive to the brush's entropy than its energy, and, thus, steepest entropy ascent causes the time evolution of the solvent density profile to deviate from the evolution of the brush's energy.

Although it may seem counter-intuitive, most of the brush's physical properties correlate well with the brush's energy.  Nevertheless, we emphasize that the path through state space does not minimize the brush's energy. If it were minimizing energy, all properties, including the solvent density profiles, would mirror the time evolution of the brush energy.

With regard to agreement between the SEAQT predictions and existing experimental data, only qualitative comparisons are possible. The predicted polystyrene density profiles in cyclohexane of Figure~\ref{Cyclo_DP} are similar to density profiles fitted to neutron reflectivity data~\cite{Karim1994} (a quantitative comparison is difficult without knowing how the neutron reflectivity data was fitted). The predicted profiles exhibit noticeable chain compression and shortening at low energy states and lengthening at higher energies, which agrees well with the experimental trends. Density profiles fitted to neutron reflectivity data obtained from polystyrene in toluene solvent~\cite{Karim1994}, however, compare less well with the SEAQT predicted profiles of Figure~\ref{Model_DP}. The SEAQT predicted profiles of a polystyrene brush in cyclohexane and in toluene are similar whereas the experimental data~\cite{Karim1994} suggests the brush density profile in toluene solvent should vary less and extend twice as far as the density profile in cyclohexane. We speculate that the lack of differentiation by SEAQT of the brush behavior in cyclohexane and toluene solvents might be attributable to an over simplistic solution model used to generate the energy eigenstructure, particularly if the greater steric complexity of toluene introduces interactions with the brush chains that are not  captured adequately by a single Flory-Huggins parameter.


On the other hand, results in~\cite{Dimitrov2007, Karim1994} for the variance of the brush height and width as a function of energy equivalent parameters, reduced temperature, and $\chi$ show similar sigmoidal behavior to the SEAQT predictions. In particular, the change in the expected normalized width for Paths I, II, and III matches the magnitude and variance of the reported equilibrium results in~\cite{Karim1994}. However, further comparisons with other experimental results are difficult. For instance, the experimental results for polystyrene interacting with cyclohexane versus those for polystyrene interacting with toluene show significant differences in extension at equivalent temperatures, with the latter showing nearly two times the extension under similar conditions. The difference in  Flory-Huggins parameter between solvents seems minor~\cite{Mark2006} in light of Equation~(\ref{FloryHuggins}) ($\approx 0.5$ for cyclohexane and $0.4$ for toluene), but the source of the difference in extension is not addressed in Karim et al~\cite{Karim1994}.

\begin{figure}
	\begin{center}
		\includegraphics[width=0.45\textwidth]{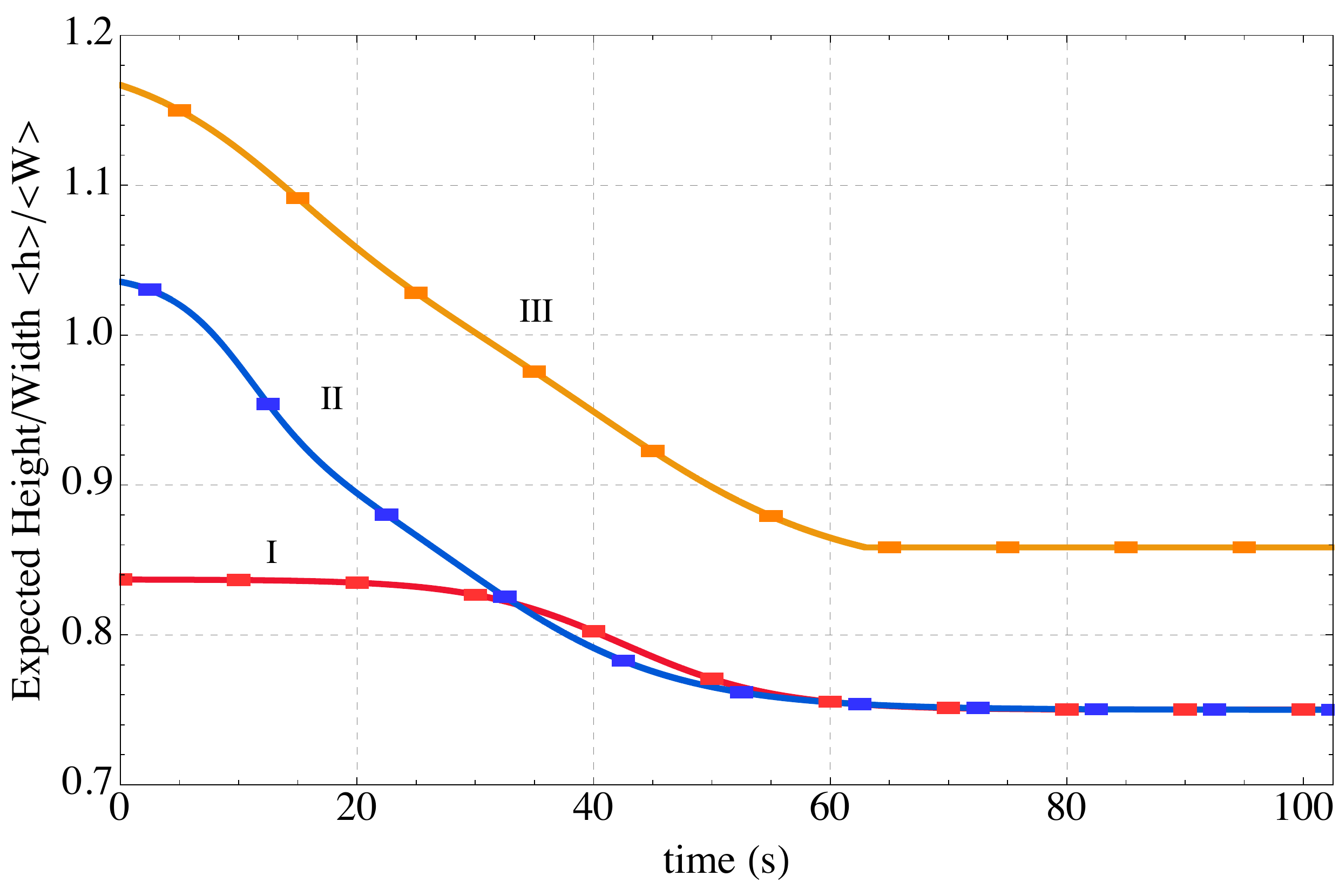}
		\caption{Ratio of the expected brush height to width in the polystyrene-cyclohexane system for Paths I, II, and III.}
		\label{H/W}
	\end{center}
\end{figure}

In contrast to some other computational models, the variations in the Flory-Huggins parameter used in this study to modify polymer-solvent interactions were intended to reflect realistic values for the energy calculations. Several computational studies~\cite{vanEck2020,He2007,Dimitrov2007,Veldscholte2020} have reported chain extensions that are qualitatively closer to experimental data, but they utilize non-physical potential parameters without explanation or justification. These studies~\cite{vanEck2020,He2007,Dimitrov2007,Veldscholte2020} also use a simplified expression for the Flory-Huggins parameter, which does not depend on temperature or local interactions, and the $\varepsilon$ interaction parameter is based on a modified Flory-Huggins parameter that varies between $1$ and $0$. However, values between $0$ and approximately $0.5$ are physically impossible. Additionally, these studies have significant disagreements with experimental data. For instance, an analysis of the brush parameters derived from published density profiles~\cite{Dimitrov2007} using Mathematica for numerical integration produces a ratio of the brush height to width for the experimental results of Karim {\em et al.}~\cite{Karim1994} of $0.7$ to $1.8$, while for Dimitrov {\em et al.}~\cite{Dimitrov2007}, the range is approximately $2.4$ to $5.2$. The analysis of the profiles developed in this study shows a ratio range of $0.74$ to $1.2$, as seen in Figure~\ref{H/W}. Although other computational models~\cite{vanEck2020,He2007,Dimitrov2007,Veldscholte2020} have shown more significant extensions by varying the Flory-Huggins parameter, the models formulated in this study were selected to adhere to realistic values for energies rather than using more simplified models.

The limited research on the temperature response of brush systems makes it difficult to identify other unaccounted-for considerations. The results seen in the literature highlight the weakness of using physical Flory-Huggins parameter values in molecular dynamic and Monte Carlo simulations. In particular, modeling the solvent on the same lattice as that of the polymer system could introduce an unnatural impediment to chain motion because of differences in migration speeds of the chain and solvent~\cite{Gupte1988,McCall1965}. In future model improvements, the solvent molecules could be placed on a separate intersecting lattice to provide the brush with greater freedom of movement while potentially improving total extension comparisons with the literature.




The concept of using an energy eigenstructure to describe the evolving conformations of polymeric systems is not unique to the work presented here. In fact, there has been increasing interest in describing polymer evolution in terms of thermodynamic landscapes, both in computational and experimental studies. However, such energy landscapes differ from the Wang-Landau formulation and are typically based upon a free energy and the natural logarithm of available conformations. The use of free energy assumes a certain entropy for given states, without considering energy level occupation. Additionally, to the authors' knowledge, there is currently no equation of motion available to generate folding paths across free energy landscapes, which hinders verification of any potential path.

Finally, our Wang-Landau and SEAQT approach offers advantages over other methods, as it overcomes the deficiencies mentioned earlier. Our derived energy eigenstructure is path-independent, meaning it is calculated without any consideration for thermodynamic equilibrium or the kinetics. Additionally, the SEAQT equation of motion uses a well-founded definition for the entropy that is associated with the probabilities of accessible states, allowing it to predict unique non-equilibrium thermodynamic paths without any \textit{a priori} assumptions of kinetic mechanisms. Furthermore, our approach eliminates the need for repeated runs, which is often required in stochastic computational approaches such as Monte Carlo simulations.

\section{Conclusions}\label{Conclusions}   
   
The conformal evolution of a polystyrene brush interacting with a cyclohexane or toluene solvent is studied utilizing an estimated path-independent energy eigenstructure with the steepest-entropy-ascent principle employed to describe the kinetic evolution of the system. Expectation structural parameters and density profiles are predicted and their time-evolution are described along three unique non-equilibrium paths. The energy eigenstructure is generated utilizing the Replica Exchange Wang-Landau method. The landscape consists of the discrete energy levels and their associated degeneracies for a model brush made up of eight polystyrene chains immersed in a solvent. The SEAQT equation of motion is then able to predict the unique thermodynamic evolution of the state of the brush system without specific details of the kinetic mechanisms involved. 

For the kinetic paths and model parameters considered here, the following conclusions are drawn:
\vspace{.2cm}
\begin{enumerate}

\item The kinetics predicted by the SEAQT equation of motion agree qualitatively with experimental density profiles from polystyrene-cyclohexane systems in the literature;  

\item The timing of the folding kinetics of the polymer brush system predicted by the the SEAQT equation of motion can be related to experimental kinetics via the SEAQT relaxation parameter;

\item Representative brush system conformations derived from state-based expectation calculations can be constructed to define the microstructural evolution along any kinetic path;

\item The farther a path lies from equilibrium, the more brush properties can be expected to vary along the path. For example, tortuosity varies two times more along non-equilibrium Path II than along non-equilibrium Path I, since the latter is comparatively closer to equilibrium; one can, thus, conclude that our non-equilibrium predictions differ significantly from the equilibrium predictions found in the literature.

\item The physical properties of the brush (radius of gyration, tortuosity, brush height and width, and polymer density profile) correlate well with energy as the brush evolves along a non-equilibrium path, but the solvent density profile is more sensitive to the brush's entropy;  

\item The choice of Flory-Huggins parameter affects the solvent distribution in the brush, but has comparatively little influence on the polymer brush density;

\item Based upon an \textit{a priori} calculation of the system energy eigenstructure, the SEAQT framework is able to go beyond current brush models to determine temperature-dependent structural evolutions.

\end{enumerate}

\section{Acknowledgements}
JM acknowledges support from the U.S.~Department of Education through the Graduate Assistance in Areas of National Need Program
(grant number P200A180016).

\bibliography{JaredsRefs_Brush.bib} 

\end{document}